\documentclass[10pt,twocolumn,twoside]{IEEEtran}
\usepackage{setspace}
\usepackage{stfloats}
\usepackage{float}
\usepackage{cite}
\usepackage{psfig}
\usepackage{subfigure}
\usepackage[dvips]{graphicx}
\usepackage{amsmath}
\usepackage{amssymb}
\usepackage{multirow}

\newcommand{\nn}{\nonumber}

\usepackage{mathtools}
\newlength{\arrow}
\newlength{\arrows}
\newcommand*{\myrightarrow}[1]{\xrightarrow{\mathmakebox[\arrow]{#1}}}
\newcommand*{\myrightarrows}[1]{\xrightarrow{\mathmakebox[\arrows]{#1}}}

\newcommand{\mysection}[1]{\vspace{-0cm}\section{#1}\vspace{-0cm}}
\newcommand{\mysubsection}[1]{\vspace{-0cm}\subsection{#1}\vspace{-0cm}}

\begin{document}
\newcounter{count1}
\newcounter{count2}
\newcounter{count3}
\newcounter{count4}
\title{Reversible Joint Hilbert and Linear Canonical Transform Without Distortion}

\author{Soo-Chang Pei,~\IEEEmembership{Fellow,~IEEE,}
and Shih-Gu Huang
\thanks{Copyright (c) 2012 IEEE. Personal use of this material is permitted. However, permission to use this material for any other purposes must be obtained from the IEEE by sending a request to pubs-permissions@ieee.org.  

This work was supported by the National Science Council, Taiwan, under Contract 98-2221-E-002-077-MY3.

S.~C. Pei is with the Department of Electrical Engineering \& Graduate Institute of Communication Engineering, National Taiwan University, Taipei 10617, Taiwan (e-mail: pei@cc.ee.ntu.edu.tw).

S.-G. Huang is with the Graduate Institute of Communication Engineering, National Taiwan University, Taipei 10617, Taiwan (e-mail: d98942023@ntu.edu.tw).
}
}

\maketitle
\begin{abstract}
Generalized analytic 
signal associated with the linear canonical transform (LCT) was proposed recently \cite{fu2008}.
However, most real signals, especially for baseband real signals, cannot be perfectly recovered from their generalized analytic signals.
Therefore, in this paper, the conventional 
Hilbert transform (HT) and analytic signal associated with the LCT 
are concerned.
To 
transform a real signal into the LCT of its HT, two integral transforms (i.e., the HT and LCT) are required.
The goal of this paper is to simplify cascades of multiple integral transforms, which may be 
the HT, analytic signal, LCT or inverse LCT.
The proposed transforms can reduce the complexity when realizing the relationships 
among the following six kinds of signals: a real signal, its HT and 
analytic signal, and the LCT of these three signals.
Most importantly, all the proposed transforms are reversible and undistorted.
Using the proposed transforms, several signal processing applications are discussed and show the advantages and flexibility over simply using the analytic signal or the LCT.
\end{abstract}

\begin{keywords}
Analytic signal, fractional Hilbert transform, generalized analytic signal, Hilbert transform, linear canonical transform
\end{keywords}

\mysection{Introduction}\label{sec:Intro}

The Hilbert transform (HT) is a linear operator connecting the real and imaginary parts of an analytic function.
The HT plays an important role in various subjects of signal processing, image processing and optics.
One of the most important subjects is the construction of analytic signals.
The analytic signal (AS) of a real-valued signal $x(t)$ is defined as
\begin{equation}\label{eq:Intro02}
x_A(t)\triangleq{\cal A}\{x(t)\}= x(t)+j\hat x(t)
\end{equation}
where $\hat x(t)$ is  the HT of $x(t)$,
\begin{equation}\label{eq:Intro03}
\hat x(t)\triangleq{\cal H} \{x(t)\}.
\end{equation}
Although $x_A(t)$ contains only non-negative frequencies of $x(t)$,
one can recover $x(t)$ from
the real part of $x_A(t)$ without any distortion due to the 
Fourier transform Hermitian property of $x(t)$.
This explains why 
analytic signals are commonly used in modulation and demodulation \cite{gabor46,feldman11}.
The analytic signal can also be expressed in terms of complex polar 
form, i.e., $x_A(t)=a(t)e^{j\phi(t)}$.
Accordingly, 
analytic signals arise 
in wide signal processing applications involving amplitude envelope and 
phase, such as phase retrieval \cite{taylor81}, instantaneous frequency estimations \cite{boashash921,boashash922}, time delay and group delay estimations \cite{marple1999,boashash2003}, quadratic time-frequency distributions \cite{boashash2003}, the Hilbert–-Huang transform \cite{huang1998,huang2005}, QRS detection from ECG \cite{benitez2001,wilson2008}, and so on.

The LCT, first introduced in \cite{collins1970,moshinsky1971}, is a parameterized general linear integral transform.
Many well-known signal processing operations, such as the Fourier transform, the fractional Fourier transform (FRFT), the Fresnel transform, and scaling and chirp multiplication operations, are the special cases of the LCT \cite{ozaktas2001,ding2001,pei2002}.
The LCT is an important tool in optics because 
the paraxial light propagation through a first-order optical system can be modeled by the LCT \cite{nazarathy1982,bastiaans1989,ozaktas2001}.
Besides, the LCT 
is very useful for filter design, radar  system  analysis, signal synthesis, time-frequency analysis, phase  reconstruction, pattern  recognition, graded index media analysis, encryption, modulation, and many other applications \cite{barshan1997,pei2000,pei2001,bastiaans2003,hennelly2005,sharma2006}.

\begin{figure*}[t]
\centering
\subfigure[] {
\centering
\includegraphics[width=1.7in]{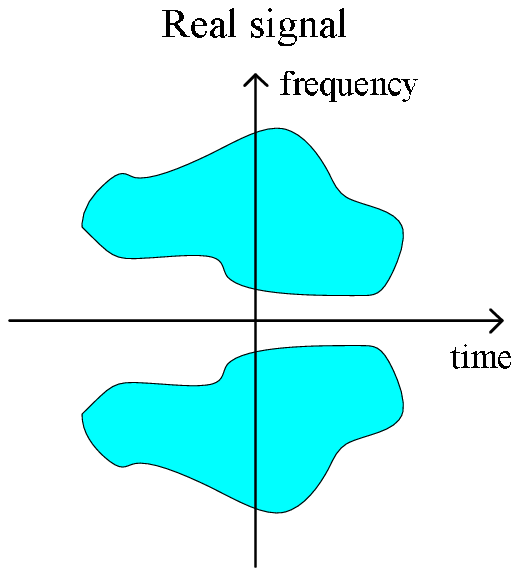}  }
\ \ \quad
\subfigure[] {
\centering
\includegraphics[width=1.7in]{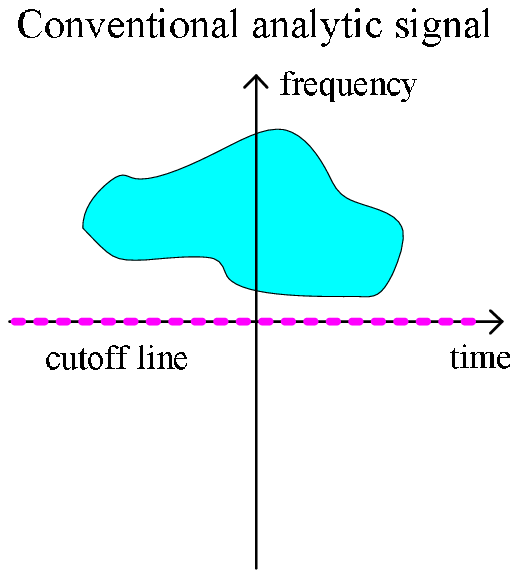}  }
\ \ \quad
\subfigure[] {
\centering
\includegraphics[width=1.7in]{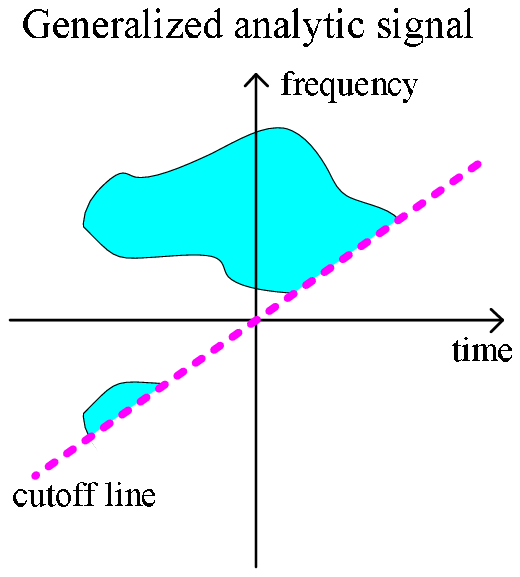}  }

\caption{
Time-frequency distributions of (a) a real signal $x(t)$, (b) conventional 
analytic signal of $x(t)$, and (c) generalized 
analytic signal of $x(t)$ using the 
PHT.
The cutoff 
lines in (b) and (c) separate the positive and negative portions 
of $x(t)$ in the Fourier domain and 
LCT domain, respectively.
}
\label{fig:FHT123}
\end{figure*}
Due to the practicality of the analytic signal over the real signal and the flexibility of the LCT over the Fourier transform, the
main goal of this paper is to derive low-complexity, reversible and undistorted transforms which combine the analytic signal and the LCT.
The 
HT and analytic signal associated with the LCT were first introduced in the generalization of the HT.
In 1996, Lohmann \emph{et al.} \cite{lohmann1996} 
introduced the fractional HT (FHT).
Instead of applying a $\pi/2$ phase shifter in the frequency domain (i.e., a sign function) as in the conventional HT, 
the FHT operates in the fractional Fourier domain using the FRFT.
The extensions of the FHT include discrete version of the FHT \cite{pei2000b,tseng2000}, factional analytic signal \cite{cusmariu2002}, and the 
generalization of the FHT \cite{tao2008}.
It has been found that the FHT is useful for image compression and edge enhancement \cite{lohmann1996,pei2000b}, and secure single-sideband (SSB) modulation \cite{tseng2000,tao2008}.
Another generalization of the HT, called generalized HT (GHT), was
proposed by Zayed \cite{zayed1998}. Instead of using the FRFT as in 
the FHT, the GHT uses a chirp function of the form $e^{-j\frac{\cot(\alpha)}{2}t^2}$.
In \cite{fu2008}, Fu and Li extended the GHT to the LCT domain, which is termed parameter $(a,b)$-Hilbert transform (PHT).
The PHT uses a chirp function of the form $e^{-j\frac{a}{2b}t^2}$, and thus is in fact equivalent to the GHT when $\frac{a}{b}=\cot(\alpha)$.
Since this paper is focused on the LCT, only the PHT is discussed in the following.

Replacing the HT $\hat x(t)$ in (\ref{eq:Intro02}) by the PHT, the resulting signal is termed generalized analytic signal (GAS).
If the parameter $a$ in the PHT is $a=0$, the PHT is 
reduced to the conventional HT, and the GAS is 
reduced to the conventional analytic signal.
If $a\neq0$, the GAS no longer contains only the non-negative components in the Fourier domain; and thus
many properties and applications of the conventional 
analytic signal do not hold for the 
GAS.
Most importantly, as $a\neq0$,
most real signals,
especially for baseband real signals, cannot be recovered from their 
GASs without distortion.
For example, consider a real signal $x(t)$ with time-frequency distribution (TFD) as shown in Fig.~\ref{fig:FHT123}(a), which is symmetric about the vertical axis due to $X(-f)=X^*(f)$.
The cutoff lines in Fig.~\ref{fig:FHT123}(b) and (c) separate the positive and negative portions of $x(t)$
in the Fourier domain and 
the LCT domain, respectively.
It is obvious that the conventional 
analytic signal shown in Fig.~\ref{fig:FHT123}(b) contains the whole information of $x(t)$, and thus can be used to reconstruct $x(t)$
perfectly.
However, for the 
GAS depicted in Fig.~\ref{fig:FHT123}(c), $x(t)$ cannot be recovered
losslessly.
As $x(t)$ has energy more concentrated in the baseband, the distortion will be greater.

When $a=0$, 
the GAS is reduced to the conventional analytic signal and irrelevant to the LCT.
When $a\neq0$, the GAS is irreversible.
Accordingly,
the \textbf{conventional analytic signal and HT} associated with the \textbf{LCT} are considered for designing reversible and undistorted transforms.
Denote ${{\cal L}^M}$ as the LCT with parameter matrix $M$,
which will be introduced in the next section, and define the $M$-LCT of an arbitrary signal $z(t)$ as ${{\cal L}^M_z}(\omega)$,
\begin{equation}\label{eq:Intro06}
{{\cal L}^M_z}(\omega)\triangleq {{\cal L}^M}\{z(t)\}.
\end{equation}
Consider a real signal $x(t)$. 
For a comprehensive understanding of the analytic signal and HT associated with the LCT, all the relationships 
among the following six kinds of signals
are investigated:
$x(t)$, $\hat x(t)$, $x_A(t)$, ${{\cal L}^M_x}(\omega)$, ${{\cal L}^M_{\hat x}}(\omega)$ and ${{\cal L}^M_{x_A}}(\omega)$.
It can be found that for some relationships, two or more integral transforms are required.
For example, to obtain ${{\cal L}^M_{x_A}}(\omega)$ from $x(t)$, two integral transforms (i.e., the 
analytic signal and LCT) are required; and to obtain ${{\cal L}^M_{\hat x}}(\omega)$ from ${{\cal L}^M_{x}}(\omega)$, three integral transforms
(i.e., the inverse LCT, HT and LCT) are used.
Therefore, the main objective of this paper is to simplify cascades of multiple integral transforms into the so-called \emph{joint transforms} 
in this paper.
Using the joint transforms to realize the relationships can reduce computational complexity.
Besides, all the joint transforms are
reversible without any distortion.
All the joint transforms are also verified by numerical simulations.
These simulations show that the numerical differences between the joint transforms and the cascades of integral transforms are down to $10^{-7}$ or less, which may be caused by numerical round-off error.

Since the joint transforms are related to the analytic signal, HT and LCT, several signal processing applications of the analytic signal, HT and LCT can be extended to the joint transforms.
For the joint transform combining the advantages of the analytic signal and the flexibility of the LCT, it can be expected that using the joint transform is preferred than simply using the analytic signal or the LCT.

This paper is organized as follows.
Section~\ref{sec:Pre} provides some useful integrals involving exponential functions.
The definitions and some properties of the HT, 
analytic signal and LCT are also introduced in this section.
In Section~\ref{sec:JT}, the definitions and derivations of all the joint transforms associated with the HT, 
analytic signal and LCT are presented.
Some simulations are 
given in Section~\ref{sec:Sim} to
verify the joint transforms and depict the advantage of them.
Several signal processing
applications of the joint transforms are discussed in Section~\ref{sec:Apc}.
Finally, conclusions are 
made in Section~\ref{sec:Con}.

\mysection{Preliminaries}\label{sec:Pre}

\mysubsection{Some Useful Integrals Involving Exponential Functions}\label{sec:Int}
It has been indicated in \cite{Gradshteyn2007} that for ${\mathop{\rm Re}\nolimits} \{ p\}  > 0$,
\begin{equation}\label{eq:Int00}
\int\limits_{ - \infty }^\infty  {{e^{ - p{t^2} \pm qt}}} \ dt = \sqrt {\frac{\pi }{p}} \ {e^{\frac{{{q^2}}}{{4p}}}}.
\end{equation}
Accordingly, the Fourier transform of a chirp function is also a chirp,
\begin{equation}\label{eq:Int04}
{\cal F}\left\{ {{e^{j\pi \gamma {t^2}}}} \right\} = {\rm{PV}}\int\limits_{ - \infty }^\infty  {{e^{j\pi \gamma {t^2} - j2\pi ft}}}\ dt = \sqrt {\frac{1}{{ - j\gamma }}} \ {e^{ - j\pi \frac{1}{\gamma }{f^2}}}.
\end{equation}
where $\sqrt {\frac{1}{{ - j\gamma }}}  = \frac{1}{{\sqrt { - j\gamma } }}$.
The symbol PV, called the \emph{Cauchy principle value}, is used to assign values to the improper integral in (\ref{eq:Int04}).
In \cite{Gradshteyn2007}, it has also been shown that for ${\mathop{\rm Re}\nolimits} \{ p\}  > 0$,
\begin{equation}\label{eq:Int05}
\int\limits_0^\infty  {{e^{ - \frac{1}{{4p}}{f^2} - qf}}} \ df = \sqrt {\pi p} \ {e^{p{q^2}}}\left[ {1 - {\rm{erf}}\left( {q\sqrt p } \right)} \right]
\end{equation}
where ${\rm erf}(x) = \frac{2}{{\sqrt \pi  }}\int_0^x {{e^{ - {t^2}}}} dt$ is the \emph{error function}.
Consider a function $G(f)$ of form
\begin{equation}\label{eq:Int08}
G(f)=2\sqrt {\frac{b}{{ - ja}}}\ u(f)\ {e^{ - j\pi \frac{b}{a}{f^2}}}
\end{equation}
where $u(f)$ denotes the unit step function. 
Based on (\ref{eq:Int05}), the inverse Fourier transform of $G(f)$ is given by
\begin{align}\label{eq:Int10}
g(t) ={\cal F}^{-1}\{G(f)\}
&={e^{j\pi \frac{a}{b}{t^2}}}\left[ {1 - {\rm{erf}}\left( { - j\sqrt { - j\pi \frac{a}{b}}\ t} \right)} \right]\nn\\
&=g_1(t)+g_2(t)
\end{align}
where $g(t)$ is also known as the \emph{Faddeeva function}
\cite{fadeeva1954} with input $\sqrt { - j\pi \frac{a}{b}}\ t$, and
\begin{align}\label{eq:LA10}
{g_1}(t) &= \ {e^{j\pi \frac{a}{b}{t^2}}}\nn\\
{g_2}(t) &=  - {e^{j\pi \frac{a}{b}{t^2}}}{\rm{erf}}\left( { - j\sqrt { - j\pi \frac{a}{b}} \ t} \right).
\end{align}
It will be shown later that ${g_1}(t)$ and ${g_2}(t)$ are widely used in the joint transforms.
Note that PV is also used in the inverse Fourier transform of (\ref{eq:Int10}).
However, throughout the rest of this paper, symbol PV is \textbf{omitted} to simplify formula expressions.

\mysubsection{Hilbert Transform and Analytic 
Signal}\label{subsec:HT}
The Hilbert transform (HT) \cite{hilbert1912,king2009} on the real line is defined as
\begin{equation}\label{eq:HT02}
\hat x(t) \triangleq {\cal H} \{x(t)\}  = x(t)*\frac{1}{{\pi t}}
\end{equation}
where $*$ denotes convolution.
Taking the Fourier transform of both sides of (\ref{eq:HT02}) with respect to $t$ yields
\begin{equation}\label{eq:HT04}
{\cal F} \{\hat x(t)\}  = {\cal F} \left\{{{\cal H} \{x(t)\} }\right\}  =  - j{\mathop{\rm sgn}} (f) \cdot X(f)
\end{equation}
where ${\rm sgn}$ denotes the sign function,
i.e., ${\mathop{\rm sgn}}(f)$ is $1$ for $f>0$, $-1$ for $f<0$ and $0$ for $f=0$.
This equation implies that applying the HT twice to 
$x(t)$ yields $-x(t)$; and thus,
the inverse HT (IHT) can be written symbolically as
\begin{equation}\label{eq:HT08}
{{\cal H}^{ - 1}} =  - {\cal H}.
\end{equation}
A list of properties, extensions and applications of the HT
have been organized in \cite{king2009,king2009v2}.

For a real signal $x(t)$, it is well known that its Fourier transform $X(f)$ is guaranteed to be Hermitian, i.e., $X(-f)=X^*(f)$.
Accordingly, the 
non-negative frequencies 
contain the whole information of $x(t)$.
Discarding the negative frequencies of $x(t)$ leads to a complex signal, $x_A(t)$, with Fourier transform given by
\begin{equation}\label{eq:HT10}
{X_A}(f) = 2u(f) \cdot X(f) = \left[ {1 + {\mathop{\rm sgn}} (f)} \right] \cdot X(f).
\end{equation}
From (\ref{eq:HT04}) and (\ref{eq:HT10}), it is obvious that $x_A(t)$ can be obtained from the HT of $x(t)$; that is,
\begin{equation}\label{eq:HT12}
x_A(t)\triangleq {\cal A} \{x(t)\}=x(t)+j{{\cal H} \{x(t)\}}.
\end{equation}
Although $x_A(t)$ is complex-valued, it occupies only half bandwidth of that of $x(t)$.
Besides, $x(t)$ can be easily recovered from $x_A(t)$ through ${\mathop{\rm Re}\nolimits} \left\{ {{x_A}(t)} \right\} = x(t)$.
Since $x_A(t)$ contains no negative frequencies, it is called an
\emph{analytic signal}.

\begin{figure*}[b]
\normalsize
\hrulefill
\setcounter{count3}{\value{equation}}
\setcounter{equation}{23}
\begin{equation}\label{eq:JT02}
\begin{array}{ccccccccccc}
 & &x(t)&\myrightarrows{\ \mathcal{A}\ }&{x_A}(t)&\myrightarrow{\ {{\mathcal{L}^M}}\ }& {\cal L}^M_{x_A}(\omega)
&\equiv&  x(t)&\myrightarrows{\ {L{A_M}}\ } &{\cal L}^M_{x_A}(\omega) \\
 & &x(t)&\myrightarrows{\ \mathcal{H}\ }&{\hat x}(t)&\myrightarrow{\ {{\mathcal{L}^M}}\ } &{\cal L}^M_{\hat x}(\omega)
&\equiv&  x(t)&\myrightarrows{\ {L{H_M}}\ } &{\cal L}^M_{\hat x}(\omega) \\
 & &{\cal L}^M_{x}(\omega) &\myrightarrows{\ {{\mathcal{L}^{M^{-1}}}}\ }&x(t)&\myrightarrow{\ \mathcal{A}\ }& {x_A}(t)
&\equiv&  {\cal L}^M_{x}(\omega)&\myrightarrows{\ AL^{-1}_M \ }& {x_A}(t) \\
 & &{\cal L}^M_{x}(\omega) &\myrightarrows{\ {{\mathcal{L}^{M^{-1}}}}\ }&x(t)&\myrightarrow{\ \mathcal{H}\ } &{\hat x}(t)
&\equiv&  {\cal L}^M_{x}(\omega)&\myrightarrows{\ HL^{-1}_M \ } & {\hat x}(t) \\
{\cal L}^M_{x}(\omega)& \myrightarrow{\ {{\mathcal{L}^{M^{-1}}}}\ }&x(t)&\myrightarrows{\ \mathcal{H}\ } &{\hat x}(t)&\myrightarrow{\ {{\mathcal{L}^M}}\ }& {\cal L}^M_{\hat x}(\omega)
&\equiv&  {\cal L}^M_{x}(\omega)&\myrightarrows{\ \left(LHL^{-1}\right)_M \ } & {\cal L}^M_{\hat x}(\omega)  \\
{\cal L}^M_{x_A}(\omega) &\myrightarrow{\ {{\mathcal{L}^{M^{-1}}}}\ }&x_A(t)&\myrightarrows{\rm{conjugation}}  &x_A^*(t)&\myrightarrow{\ {{\mathcal{L}^M}}\ }& {\cal L}^M_{x_A^*}(\omega)
&\equiv&  {\cal L}^M_{x_A}(\omega)&\myrightarrows{\ \left(LcL^{-1}\right)_M \ }   &{\cal L}^M_{x_A^*}(\omega)
\end{array}
\end{equation}
\setcounter{count4}{\value{equation}}
\setcounter{equation}{\value{count3}}
\end{figure*}

\mysubsection{Linear Canonical Transform}\label{subsec:LCT}
In this paper, the definition of the linear canonical transform (LCT) with four parameters \cite{ozaktas2001,pei2011} is adopted,
\begin{align}\label{eq:LCT02}
{{\cal L}^M_x}(\omega)&\triangleq {{\cal L}^M}\{x(t)\}\nn\\
&= \left\{
  \begin{array}{l l}
\sqrt {\dfrac{1}{{jb}}}\ {e^{j\pi \frac{d}{b}{\omega ^2}}} \int\limits_{ - \infty }^\infty  {{e^{j\pi \frac{a}{b}{t^2}}}{e^{ - j2\pi \frac{\omega }{b}t}}} \ x(t)\ dt,& \ b \ne 0\\
{\sqrt d \ {e^{j\pi \,cd\,{\omega ^2}}}x(d\omega ),}& \ b = 0
  \end{array} \right.
\end{align}
where ${{\cal L}^M}$ is the LCT operator with
\begin{equation}\label{eq:LCT04}
M = (a,b,c,d) = \begin{bmatrix}
a\ &b\\
c\ &d
\end{bmatrix}\quad {\rm and} \quad ad - bc = 1.
\end{equation}
Some properties of the LCT, which will be used later, are listed below:
\begin{itemize}
\item Conjugation property \cite{pei2002}\\
The conjugate of the $(a,b,c,d)$-LCT of $z(t)$ is equivalent to the $(a,-b,-c,d)$-LCT of the conjugate of $z(t)$, i.e.,
\begin{equation}\label{eq:LCT08}
{\left[ {{\cal L}_z^{(a,b,c,d)}(\omega )} \right]^*} = {\cal L}_{{z^*}}^{(a, - b, - c,d)}(\omega ).
\end{equation}

\item Inverse LCT (ILCT) \cite{ozaktas2001,pei2002}\\
The inverse of the $M$-LCT is given by
\begin{equation}\label{eq:LCT12}
{\left[ {{{\cal L}^M}} \right]^{ - 1}} = {{\cal L}^{{M^{ - 1}}}}
\end{equation}
where $M^{ - 1}=(d,-b,-c,a)$ is the inverse of matrix $M$.

\item Equivalent 
expressions of the LCT ($b\neq0$)\\
If $a=0, b\neq0$, the LCT in (\ref{eq:LCT02}) can be rewritten as
\begin{align}\label{eq:LCT14}
{{\cal L}^M_x}(\omega)&=
\sqrt {\dfrac{1}{{jb}}}\ {e^{j\pi \frac{d}{b}{\omega ^2}}}\int\limits_{ - \infty }^\infty  {{e^{ - j2\pi \frac{\omega }{b}t}}} \ x(t)\ dt\\
&={\sqrt {\frac{1}{{jb}}} \ {e^{j\pi \frac{d}{b}{\omega ^2}}}X\left( {\frac{\omega }{b}} \right)}.\label{eq:LCT16}
\end{align}
For $a\neq0, b\neq0$, the LCT in (\ref{eq:LCT02}) can 
be expressed as three 
\textbf{equivalent}
forms.
For ease of distinguishing these LCT expressions, the original definition of the LCT with $b\neq0$ given in (\ref{eq:LCT02}) is called \textbf{LCT Form I}, and the three 
equivalent forms are called 
LCT Forms II, III and IV, respectively:
\begin{align}
&{\rm \textbf{LCT Form II}}  \nn\\
&\ {{\cal L}^M_x}(\omega)=\sqrt {\frac{1}{{jb}}} \ {e^{j\pi \frac{c}{a}{\omega ^2}}}{\left[ {x(t) *{e^{j\pi \frac{a}{b}{t^2}}}} \right]_{\ t = \frac{\omega }{a}}} \label{eq:LCT18} \\
&{\rm \textbf{LCT Form III}}  \nn\\
&\ {{\cal L}^M_x}(\omega)= |d|\sqrt {\frac{1}{{jb}}} \left( {x(d\omega ){e^{j\pi cd{\omega ^2}}}} \right) * {e^{j\pi \frac{d}{b}{\omega ^2}}} \label{eq:LCT20} \\
&{\rm \textbf{LCT Form IV}}  \nn\\
&\ {{\cal L}^M_x}(\omega)=\sqrt {\frac{1}{{jb}}} \sqrt {\frac{b}{{ - ja}}} {\ e^{j\pi \frac{c}{a}{\omega ^2}}}\nn\\
&\quad\qquad\qquad\qquad\cdot\int\limits_{ - \infty }^\infty  {X(f)\ {e^{ - j\pi \frac{b}{a}{f^2}}}{e^{j2\pi \frac{\omega }{a}f}}} \ df  \label{eq:LCT22}
\end{align}
where 
${e^{j\pi \frac{a}{b}{t^2}}}$ in (\ref{eq:LCT18}) has been defined as $g_1(t)$ in (\ref{eq:LA10}).
Note that $\sqrt {\frac{1}{{jb}}} \sqrt {\frac{b}{{ - ja}}} = \sqrt {\frac{1}{a}}$ is in general \textbf{not true} because $a,b$ may be negative.
The derivations of 
(\ref{eq:LCT18})-(\ref{eq:LCT22}) are 
presented in Appendix~\ref{App:AF}.

\end{itemize}

\begin{figure}[t]
\centering
\includegraphics[width=0.9\columnwidth]{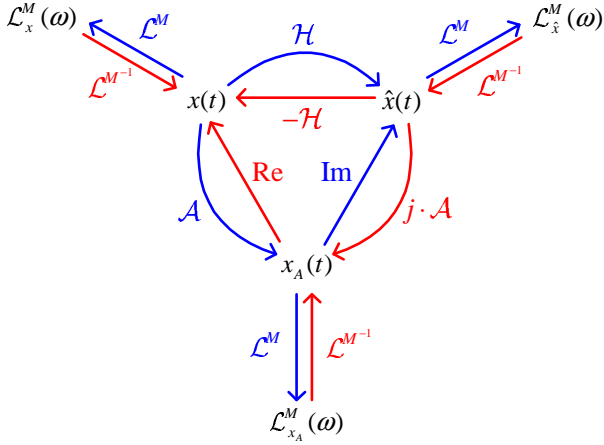}
\caption{
Relationships among $x(t)$, $\hat x(t)$, $x_A(t)$, ${{\cal L}^M_x}(\omega)$, ${{\cal L}^M_{\hat x}}(\omega)$, and ${{\cal L}^M_{x_A}}(\omega)$ 
(denote a real signal, HT of $x(t)$, 
analytic signal of $x(t)$, $M$-LCT of $x(t)$, $M$-LCT of $\hat x(t)$, and $M$-LCT of $x_A(t)$,
respectively).
Symbols $\cal H$, $\cal A$, ${\cal L}^M$, ${\cal L}^{M^{-1}}$, Re and Im denote the HT, 
analytic signal, LCT, ILCT, real part and imaginary part, respectively.
$j=\sqrt{-1}$.
}
\label{fig:JT1}
\end{figure}

\mysection{Joint Transforms Associated With the HT, 
Analytic Signal and LCT}\label{sec:JT}
In order to benefit from the advantages
of the analytic signal and the flexibility of
the LCT, we want to derive low-complexity, reversible and undistorted transforms which combine the analytic signal (or the HT) and the LCT.
As discussed in the introduction,
a kind of analytic signal associated with the LCT is the generalized analytic signal (GAS) \cite{fu2008}.
However, when the parameter $a$ in the GAS is $a\neq0$, the GAS is irreversible.
When $a=0$, 
the GAS is reduced to the conventional analytic signal and irrelevant to the LCT.
Therefore, the conventional 
analytic signal and HT associated with the LCT 
are considered.

For a comprehensive understanding of the analytic signal and HT associated with the LCT, all the
relationships among the following six kinds of signals are examined and illustrated in Fig.~\ref{fig:JT1}: $x(t)$, $\hat x(t)$, $x_A(t)$, ${{\cal L}^M_x}(\omega)$, ${{\cal L}^M_{\hat x}}(\omega)$, and ${{\cal L}^M_{x_A}}(\omega)$ 
(denote a real signal, HT of $x(t)$, 
analytic signal of $x(t)$, $M$-LCT of $x(t)$, $M$-LCT of $\hat x(t)$, and $M$-LCT of $x_A(t)$,
respectively).
Using the operators HT, 
analytic signal, LCT, ILCT, real part, and imaginary part 
(denoted by $\cal H$, $\cal A$, ${\cal L}^M$, ${\cal L}^{M^{-1}}$, Re, and Im,
respectively), one can easily transform one of the six signals into another without any information loss.
However, it can be found that two or three integral transforms are required for the relationships between $x(t)$ and ${{\cal L}^M_{\hat x}}(\omega)$, between ${{\cal L}^M_x}(\omega)$ and  ${{\cal L}^M_{x_A}}(\omega)$, and so on.
Therefore, in this section, the cascades of multiple integral transforms are simplified to our so-called \emph{joint transforms}.

All possible joint transforms regarding these six kinds of signals include 
\textbf{joint LCT-AS} ($LA_M$), \textbf{joint LCT-HT} ($LH_M$), 
\textbf{joint AS-ILCT} ($AL^{-1}_M$), \textbf{joint HT-ILCT} ($HL^{-1}_M$), \textbf{joint LCT-HT-ILCT} ($\left(LHL^{-1}\right)_M$) and 
\textbf{joint LCT-conjugation-ILCT} ($\left(LcL^{-1}\right)_M$).
The reader is reminded that \textbf{AS} is abbreviated from analytic signal.
\setcounter{equation}{\value{count4}}
The definitions of these joint transforms are listed in (\ref{eq:JT02}),
where ${\cal L}^M_{x_A^*}(\omega)$ is used in ${\cal L}^M_{x}(\omega)=\frac{1}{2}\left({\cal L}^M_{x_A}(\omega)+{\cal L}^M_{x_A^*}(\omega)\right)$ and ${\cal L}^M_{\hat x}(\omega)=\frac{1}{2j}\left({\cal L}^M_{x_A}(\omega)-{\cal L}^M_{x_A^*}(\omega)\right)$.
Use the joint LCT-AS ($L{A_M}$) as an example.
In order to generate ${\cal L}^M_{x_A}(\omega)$ from $x(t)$, conventionally, we have to find the analytic signal ($\cal A$) of $x(t)$
first, and then apply the LCT (${\cal L}^M$) to $x_A(t)$; that is, ${\cal L}^M_{x_A}(\omega)={{\cal L}^M}\left\{{\cal A}\{x(t)\}\right\}$.
In our method, ${\cal L}^M_{x_A}(\omega)$ is obtained directly from $x(t)$ by ${\cal L}^M_{x_A}(\omega)=L{A_M}\left\{ {x(t)} \right\}$, and the 
intermediate $x_A(t)$ is not generated.
All the relationships involving multiple integral transforms can be 
equivalently carried out by one of the above joint transforms.
The joint LCT-AS and joint LCT-HT can be deemed as the generalizations of the analytic signal and the HT, respectively.
As $(a,b,c,d)=(1,0,0,1)$ where the LCT is equivalent to identity operator, the joint LCT-AS is 
reduced to the conventional analytic signal, and the joint LCT-HT is 
reduced to the conventional HT.
In the following, the derivations of the joint transforms are presented.
The case of the LCT/ILCT with $b=0$ is ignored since it is 
simply a time-scaled version of $x(t)$ multiplied by a linear chirp, i.e., ${{\cal L}^M}\{x(t)\}=\sqrt d \ {e^{j\pi \,cd\,{\omega ^2}}}x(d\omega)$.

\mysubsection{Joint LCT-AS ($b\neq0$)}\label{subsec:LA}
The joint LCT-AS, denoted by $LA_M$, transforms $x(t)$ into ${\cal L}^M_{x_A}(\omega)$.
It is equivalent to calculating the analytic signal (AS) $x_A(t)$ of $x(t)$ first and then generating ${\cal L}^M_{x_A}(\omega)$ from $x_A(t)$ by the LCT,
\begin{equation}\label{eq:LA00}
{\cal L}^M_{x_A}(\omega)=L{A_M}\left\{ {x(t)} \right\} \triangleq {{\cal L}^M}\left\{{\cal A}\{x(t)\}\right\}.
\end{equation}
Alternatively, the relationship between $x(t)$ and ${\cal L}^M_{x_A}(\omega)$ can be expressed as
${\cal L}^M_{x_A}(\omega)={{\cal L}^M}\left\{x_A(t)\right\}$, where
the Fourier transform of $x_A(t)$ is $X_A(f)={\cal F}\{x(t)\}\cdot 2u(f)$.
Accordingly, if $a=0$, from (\ref{eq:LCT16}), the joint LCT-AS is give by
\begin{align}\label{eq:LA015}
&L{A_M}\{x(t)\}
=\sqrt {\frac{1}{{jb}}} \ {e^{j\pi \frac{d}{b}{\omega ^2}}}X\left( {\frac{\omega }{b}} \right) \cdot 2u\left( {\frac{\omega }{b}} \right)\\
&\quad=\sqrt {\frac{1}{{jb}}} \ {e^{j\pi \frac{d}{b}{\omega ^2}}}\left(\int\limits_{ - \infty }^\infty  {x(t)\ {e^{ - j2\pi \frac{\omega }{b}t}}} \ dt \right)\cdot 2u\left( {\frac{\omega }{b}} \right).\label{eq:LA01}
\end{align}
If $a\neq0$, based on the LCT Form IV in (\ref{eq:LCT22}), it follows that
\begin{align}\label{eq:LA02}
L{A_M}\{x(t)\}
&=\sqrt {\frac{1}{{jb}}} \sqrt {\frac{b}{{ - ja}}} {\ e^{j\pi \frac{c}{a}{\omega ^2}}} \nn\\
&\ \quad \cdot\int\limits_{ - \infty }^\infty  {X(f)\cdot 2u(f)\ {e^{ - j\pi \frac{b}{a}{f^2}}}{e^{j2\pi \frac{\omega }{a}f}}} \ df.
\end{align}
According to (\ref{eq:Int08}) and (\ref{eq:Int10}), formula (\ref{eq:LA02}) can be rewritten as
\begin{equation}\label{eq:LA08}
L{A_M}\{x(t)\}
=\sqrt {\frac{1}{{jb}}} {e^{j\pi \frac{c}{a}{\omega ^2}}}{\left[ {x(t) * \left( {{g_1}(t) + {g_2}(t)} \right)} \right]_{t = \frac{\omega }{a}}}
\end{equation}
where $g_1(t)$ and $g_2(t)$ are defined in (\ref{eq:LA10}).
The joint LCT-AS 
is also used in the transformation from ${\hat x}(t)$ to ${\cal L}^M_{x_A}(\omega)$.

\mysubsection{Joint LCT-HT ($b\neq0$)}\label{subsec:LH}
The joint LCT-HT, denoted by $LH_M$, transforms $x(t)$ into ${\cal L}^M_{\hat x}(\omega)$.
It is equivalent to transforming $x(t)$ into $\hat x(t)$ by the HT first and then transforming $\hat x(t)$ into ${\cal L}^M_{\hat x}(\omega)$ by the LCT,
\begin{equation}\label{eq:LH00}
{\cal L}^M_{\hat x}(\omega)=L{H_M}\{x(t)\}\triangleq {{\cal L}^M}\left\{{\cal H}\{x(t)\}\right\}.
\end{equation}
The joint LCT-HT can be derived from the following alternative relationship:
\begin{align}\label{eq:LH01}
{\cal L}^M_{\hat x}(\omega)&= {{\cal L}^M}\left\{ - j\left[ x_A(t) -x(t)\right] \right\} \nn\\
&=- j\left[ {LA_M\{x(t)\} - {{\cal L}^M}\{x(t)\}} \right].
\end{align}
Here, (\ref{eq:LCT14}) and the LCT Form II in (\ref{eq:LCT18}) are adopted for ${{\cal L}^M}\{x(t)\}$, 
while $LA_M\{x(t)\}$ has been derived in (\ref{eq:LA01}) and (\ref{eq:LA08}).
Therefore, the joint LCT-HT is given by
\begin{align}\label{eq:LH06}
&L{H_M}\{x(t)\}\nn\\
&= \left\{
  \begin{array}{l l}
-j\sqrt {\frac{1}{{jb}}} \ {e^{j\pi \frac{c}{a}{\omega ^2}}}{\left[ {x(t) *  {{g_2}(t)} } \right]_{\ t = \frac{\omega }{a}}},& \ a \ne 0\\
-j\sqrt {\frac{1}{{jb}}} \ {e^{j\pi \frac{d}{b}{\omega ^2}}}\int\limits_{ - \infty }^\infty  {x(t)\ {e^{ - j2\pi \frac{\omega }{b}t}}} \ dt & \\
\quad\qquad\qquad\qquad\qquad\qquad\qquad\cdot {\rm sgn}\left( {\frac{\omega }{b}} \right),& \ a = 0
  \end{array} \right.
\end{align}
where $g_2(t)$ is given in (\ref{eq:LA10}).
The joint LCT-HT is also applied to the transformation from ${\hat x}(t)$ to ${\cal L}^M_{x}(\omega)$.

\mysubsection{Joint AS-ILCT ($b\neq0$)}\label{subsec:AI}
The joint AS-ILCT, denoted by $AL^{-1}_M$, transforms ${\cal L}^M_x(\omega)$ into $x_A(t)$. 
It is equivalent to transforming ${\cal L}^M_x(\omega)$ into $x(t)$ by the ILCT first and then calculating the analytic signal (AS) $x_A(t)$ of $x(t)$,
\begin{equation}\label{eq:AI00}
x_A(t)=AL_M^{ - 1}\left\{ {\cal L}^M_x(\omega) \right\}\triangleq {\cal A}\left\{{{\cal L}^{{M^{ - 1}}}}\left\{ {\cal L}^M_x(\omega) \right\}\right\}.
\end{equation}
The joint AS-ILCT  can also be expressed as
\begin{equation}\label{eq:AI005}
AL_M^{ - 1}\left\{ {\cal L}^M_x(\omega) \right\}=\int\limits_{ - \infty }^\infty  {X(f) \cdot 2u(f)} \ {e^{j2\pi ft}}df
\end{equation}
where $X(f)$ can be determined by ${\cal L}^M_x(\omega)$ according to (\ref{eq:LCT16}) and (\ref{eq:LCT22}).
For $a=0$, formula (\ref{eq:LCT16}) implies that
\begin{equation}\label{eq:AI01}
X\left( {\frac{\omega }{b}} \right)= \sqrt {jb}\ {e^{-j\pi \frac{d}{b}{\omega ^2}}}{{\cal L}^M_x}(\omega).
\end{equation}
Substituting (\ref{eq:AI01}) into (\ref{eq:AI005}) results in
\begin{align}\label{eq:AI02}
&AL^{-1}_M\{{\cal L}^M_x(\omega)\}\nn\\
&\quad=\sqrt {\frac{1}{{ - jb}}} \int\limits_{ - \infty }^\infty  {{\cal L}_x^M(\omega )\cdot2u\left( {\frac{\omega }{b}} \right)} \ {e^{ - j\pi \frac{d}{b}{\omega ^2}}}{e^{j2\pi \frac{t}{b}\omega }}d\omega .
\end{align}
For $a\neq0$, recall LCT Form IV in (\ref{eq:LCT22}) with $\frac{\omega}{a}=\nu$, and then $X(f)$ is given by
\begin{equation}\label{eq:AI04}
X(f)=
\sqrt {jb} \sqrt {\frac{{ - ja}}{b}} {e^{j\pi \frac{b}{a}{f^2}}}\int\limits_{ - \infty }^\infty  {{\cal L}^M_x(a\nu){e^{ - j\pi ac{\nu ^2}}}{e^{ - j2\pi \nu f}}} d\nu.
\end{equation}
Again, substituting (\ref{eq:AI04}) into (\ref{eq:AI005}), it follows that
\begin{align}\label{eq:AI06}
AL^{-1}_M\{{\cal L}^M_x(\omega)\}
&= \sqrt {jb} \left| {\frac{a}{b}} \right|\int\limits_{ - \infty }^\infty{{\cal L}^M_x(a\nu)}\ {e^{ - j\pi ac{\nu ^2}}}\nn\\
&\ \ \qquad\cdot \left(\int\limits_{ - \infty }^\infty  G^*(f){e^{j2\pi \left( {t - \nu } \right)f}}df\right)   d\nu
\end{align}
where $G(f)$ has been defined in (\ref{eq:Int08}).
From (\ref{eq:Int10}), the inverse Fourier transform of $G^*(f)$ is $g_1^*(-t)+g_2^*(-t)$.
From the definitions in (\ref{eq:LA10}), it 
is obvious that $g_1^*(-t)=g_1^*(t)$,
and $g_2^*(-t)=-g_2^*(t)$ since ${\rm{erf}}\left( { - z} \right) =  - {\rm{erf}}\left( z \right)$ for any complex number $z$.
Therefore, (\ref{eq:AI06}) can further 
simplify to
\begin{align}\label{eq:AI12}
&AL_M^{ - 1}\left\{ {\cal L}^M_x(\omega) \right\}\nn\\
&= \left| a \right|\sqrt {\frac{1}{{ - jb}}} \left( {{{\cal L}^M_x(at)}\ {e^{ - j\pi ac{t^2}}}} \right)*\left( {g_1^*(t) - g_2^*(t)} \right).
\end{align}
When realizing the transformation from ${\cal L}^M_{\hat x}(\omega)$ to $x_A(t)$, the joint AS-ILCT can also be adopted.

\begin{figure*}[t]
\centering
\includegraphics[width=0.9\columnwidth]{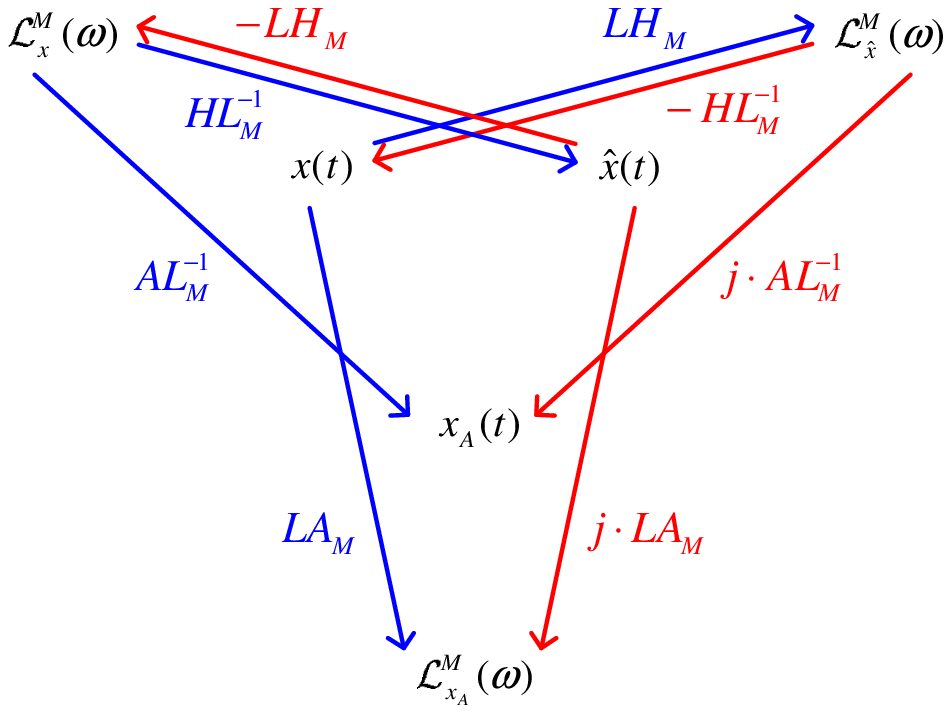}
\quad\qquad
\includegraphics[width=0.9\columnwidth]{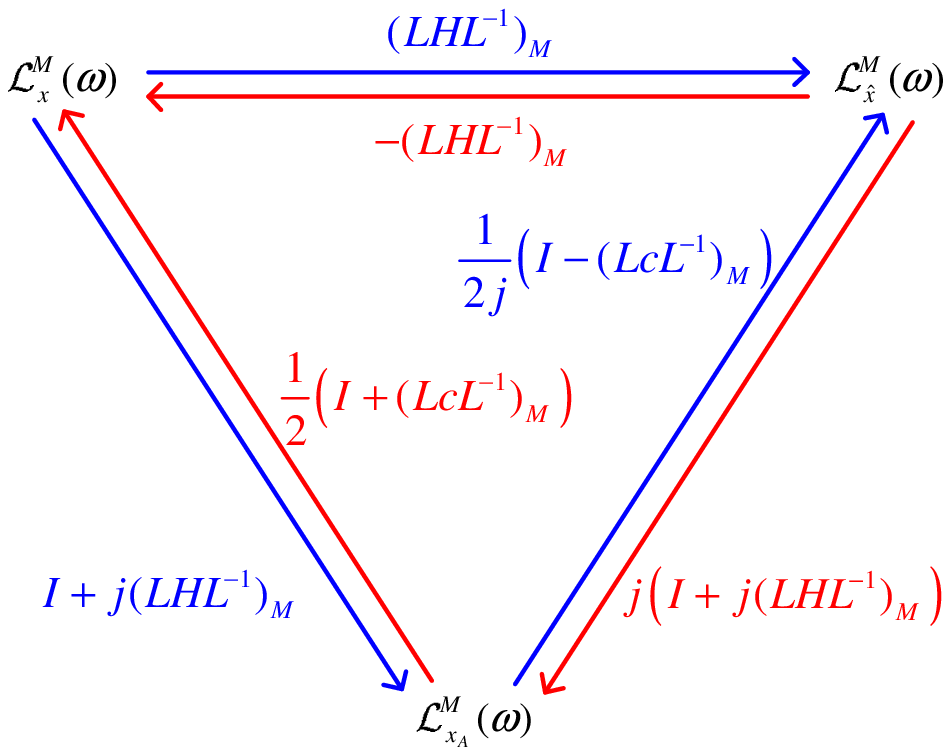}
\caption{
Relationships which can be realized by the joint transforms, including the joint LCT-AS ($LA_M$), joint LCT-HT ($LH_M$), joint AS-ILCT ($AL^{-1}_M$), joint HT-ILCT ($HL^{-1}_M$), joint LCT-HT-ILCT ($\left(LHL^{-1}\right)_M$) and joint LCT-conjugation-ILCT ($\left(LcL^{-1}\right)_M$).
$I$ is the identity operator and $j=\sqrt{-1}$.
}
\label{fig:JT23}
\end{figure*}

\mysubsection{Joint HT-ILCT ($b\neq0$)}\label{subsec:HI}
The joint HT-ILCT, denoted by $HL^{-1}_M$, transforms ${\cal L}^M_x(\omega)$ into ${\hat x}(t)$.
It is equivalent to transforming ${\cal L}^M_x(\omega)$ into $x(t)$ by the ILCT first and then calculating the HT $\hat x(t)$,
\begin{equation}\label{eq:HI00}
{\hat x}(t)=HL_M^{ - 1}\left\{ {\cal L}^M_x(\omega) \right\} \triangleq{\cal H}\left\{{{\cal L}^{{M^{ - 1}}}}\left\{ {\cal L}^M_x(\omega) \right\}\right\}.
\end{equation}
Alternatively, the following relationship is used,
\begin{align}\label{eq:HI01}
{\hat x}(t)&=  - j\left[ {{x_A}(t) - x(t)} \right]\nn\\
&=  - j\left[ {AL_M^{ - 1}\{ {\cal L}_x^M(\omega )\}  - {{\cal L}^{{M^{ - 1}}}}\left\{ {{\cal L}_x^M(\omega )} \right\}} \right].
\end{align}
For $a=0$, $AL_M^{ - 1}$ has been given in (\ref{eq:AI02}), while ${{\cal L}^{{M^{ - 1}}}}$ can be obtained by substituting $M^{-1}=(d,-b,-c,0)$ for $M=(a,b,c,d)$ in (\ref{eq:LCT02}).
Therefore, for $a=0$, the joint HT-ILCT is given by
\begin{align}\label{eq:HI04}
HL_M^{ - 1}\left\{ {{\cal L}_x^M(\omega )} \right\}
&=  - j\sqrt {\frac{1}{{ - jb}}} \int\limits_{ - \infty }^\infty  {\cal L}_x^M(\omega )\nn\\
&\qquad\cdot{\mathop{\rm sgn}} \left( {\frac{\omega }{b}} \right) \ {e^{ - j\pi \frac{d}{b}{\omega ^2}}}{e^{j2\pi \frac{t}{b}\omega }}d\omega.
\end{align}
Similarly, for $a\neq0$, the $AL_M^{ - 1}$ in (\ref{eq:AI12}) and the LCT Form III in (\ref{eq:LCT20}) with $(a,b,c,d)$ replaced by $(d,-b,-c,a)$ 
lead to
\begin{equation}\label{eq:HI08}
HL_M^{ - 1}\left\{ {{\cal L}_x^M(\omega )} \right\}
=j|a|\sqrt {\frac{1}{{ - jb}}} \left( {{\cal L}_x^M(at){e^{ - j\pi ac{t^2}}}} \right)*g_2^*(t)
\end{equation}
where $g_2(t)$ is given in (\ref{eq:LA10}).
The negative joint HT-ILCT, i.e., $-HL^{-1}_M$, can be used to transform ${\cal L}^M_{\hat x}(\omega)$ into $x(t)$.

\begin{figure*}[t]
\centering
\includegraphics[width=0.94\linewidth]{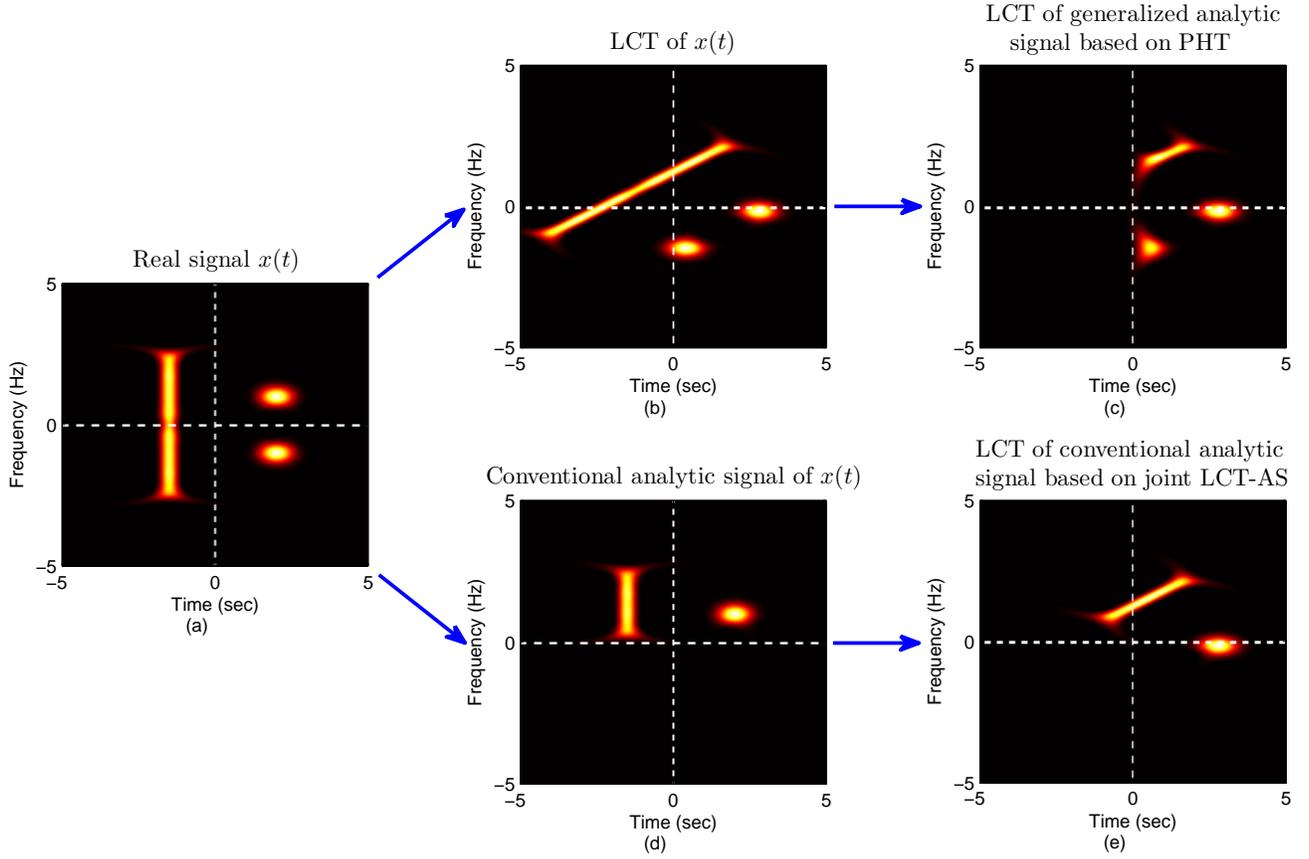}
\vspace*{-0.2cm}
\caption{
For $M=(a,b,c,d)=(0.8,1.2,-0.4,0.65)$, time-frequency distributions of (a) a baseband real signal $x(t)$;
(b) the LCT of $x(t)$ (i.e., ${{\cal L}^M_x}(\omega)$);
(c) the LCT of the generalized analytic signal based on PHT;
(d) the conventional analytic signal of $x(t)$ (i.e., $x_A(t)$); and
(e) the LCT of $x_A(t)$ (i.e., ${{\cal L}^M_{x_A}}(\omega)$) calculated directly from $x(t)$ by the joint LCT-AS.
The signal in (c) is equivalent to suppressing the negative part of ${{\cal L}^M_x}(\omega)$ in (b), and cannot be used to recover $x(t)$ without distortion because of significant information loss.
In contrast, the signal in (e) generated by the joint LCT-AS contains the whole information of $x(t)$.}
\label{fig:PHT}
\vspace*{-0.3cm}\end{figure*}

\begin{figure*}[!b]
\normalsize
\hrulefill
\setcounter{count1}{\value{equation}}
\setcounter{equation}{49}
\begin{align}\label{eq:LcI04}
\left(LcL^{-1}\right)_M\left\{ {{\cal L}_{x_A}^M(\omega )} \right\} &= {\left[ {\sqrt {\frac{1}{{ - jb}}} \ {e^{ - j\pi \frac{d}{b}{\omega ^2}}}\int\limits_{ - \infty }^\infty  {{e^{ - j\pi \frac{a}{b}{t^2}}}{e^{j2\pi \frac{\omega }{b}t}}} \
{\cal L}^{(d, - b, - c,a)}\left\{ {{\cal L}_{x_A}^M(\eta )} \right\}dt} \right]^*}\nn\\
 &= {\left[ {\frac{1}{{ - jb}}{e^{ - j\pi \frac{d}{b}{\omega ^2}}}\int\limits_{ - \infty }^\infty  {{\cal L}_{x_A}^M(\eta )\ {e^{ - j\pi \frac{d}{b}{\eta ^2}}}} \ \left( {\int\limits_{ - \infty }^\infty  {{e^{ - j\pi \frac{{2a}}{b}{t^2}}}{e^{j2\pi \frac{{\omega  + \eta }}{b}t}}} \ \ dt} \right)d\eta } \right]^*}.
\end{align}
\setcounter{count2}{\value{equation}}
\setcounter{equation}{\value{count1}}
\end{figure*}

\mysubsection{Joint LCT-HT-ILCT ($b\neq0$)}\label{subsec:LHI}
The joint LCT-HT-ILCT, denoted by $\left(LHL^{-1}\right)_M$, transforms ${\cal L}^M_x(\omega)$ into ${\cal L}^M_{\hat x}(\omega)$.
It is equivalent to transforming ${\cal L}^M_x(\omega)$ into $x(t)$ by the ILCT first, then calculating the HT $\hat x(t)$, and finally transforming $\hat x(t)$ into ${\cal L}^M_{\hat x}(\omega)$ by the LCT,
\begin{align}\label{eq:LHI00}
{\cal L}^M_{\hat x}(\omega)&=\left(LHL^{-1}\right)_M\left\{ {\cal L}^M_x(\omega) \right\}\nn\\
&\triangleq{\cal L}^M\left\{{\cal H}\left\{{{\cal L}^{{M^{ - 1}}}}\left\{ {\cal L}^M_x(\omega) \right\}\right\}\right\}.
\end{align}
Since ${\cal L}_{\hat x}^M(\omega ) =  - j\left[ {\cal L}_{{x_A}}^M(\omega ) - {\cal L}_x^M(\omega ) \right]$, now the problem is how to determine ${\cal L}_{{x_A}}^M(\omega )$ from ${\cal L}^M_x(\omega)$.
It has been shown in (\ref{eq:LA015}) and (\ref{eq:LA02}) that ${\cal L}_{{x_A}}^M(\omega )$ can be expressed in terms of $X(f)$,
and (\ref{eq:AI01}) and (\ref{eq:AI04}) show that  $X(f)$ can be expressed in terms of ${\cal L}^M_x(\omega)$.
Accordingly, for $a=0$,  substituting  (\ref{eq:AI01})  into  (\ref{eq:LA015}) results in
\begin{align}\label{eq:LHI06}
{\cal L}_{{x_A}}^M(\omega )
={\cal L}_x^M(\omega )\cdot2u\left( {\frac{\omega }{b}} \right).
\end{align}
For $a\neq0$, substituting  (\ref{eq:AI04})  into  (\ref{eq:LA02}) leads to
\begin{align}\label{eq:LHI08}
&{\cal L}_{{x_A}}^M(\omega )\nn\\
& = {e^{j\pi \frac{c}{a}{\omega ^2}}}\hspace{-5pt}\int\limits_{ - \infty }^\infty  {{\cal L}_x^M(a\nu ){e^{ - j\pi ac{\nu ^2}}}{\int\limits_{ - \infty }^\infty  {2u(f){e^{j2\pi \left( {\frac{\omega }{a} - \nu } \right)f}}} df} \ } d\nu \nn\\
& = {\cal L}_x^M(\omega ) + j{\mathop{\rm sgn}} (a){e^{j\pi \frac{c}{a}{\omega ^2}}}\hspace{-5pt}\int\limits_{ - \infty }^\infty  {{\cal L}_x^M(\eta ){e^{ - j\pi \frac{c}{a}{\eta ^2}}}\frac{1}{{\pi \left( {\omega  - \eta } \right)}}} d\eta .
\end{align}
As ${\cal L}_{\hat x}^M(\omega )$ can be 
determined from (\ref{eq:LHI06}) and (\ref{eq:LHI08}), 
we have
\begin{align}\label{eq:LHI10}
&\left(LHL^{-1}\right)_M\left\{ {{\cal L}_x^M(\omega )} \right\}\nn\\
&= \left\{
  \begin{array}{l l}
{\mathop{\rm sgn}} (a)\ {e^{j\pi \frac{c}{a}{\omega ^2}}}\int\limits_{ - \infty }^\infty  {{\cal L}_x^M(\eta )\ {e^{ - j\pi \frac{c}{a}{\eta ^2}}}\frac{1}{{\pi \left( {\omega  - \eta } \right)}}\ } d\eta ,& a \ne 0\\
-j{\cal L}_x^M(\omega )\cdot{\rm sgn}\left( {\frac{\omega }{b}} \right) ,& a = 0
  \end{array} \right..
\end{align}
Also, it is apparent that $-\left(LHL^{-1}\right)_M$ can transform ${\cal L}^M_{\hat x}(\omega)$ into ${\cal L}^M_x(\omega)$.

\begin{figure*}[t]
\addtolength{\subfigcapskip}{-7pt}
\centering{
\subfigure[] {
\centering
\includegraphics[width=0.95\columnwidth]{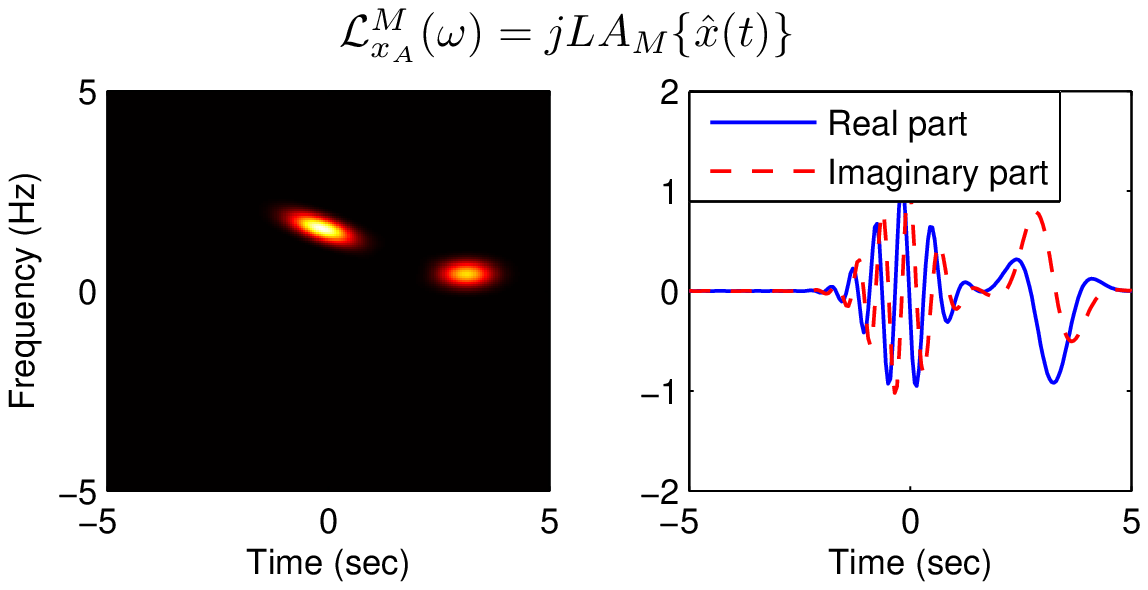}}
\quad
\subfigure[] {
\centering
\includegraphics[width=0.95\columnwidth]{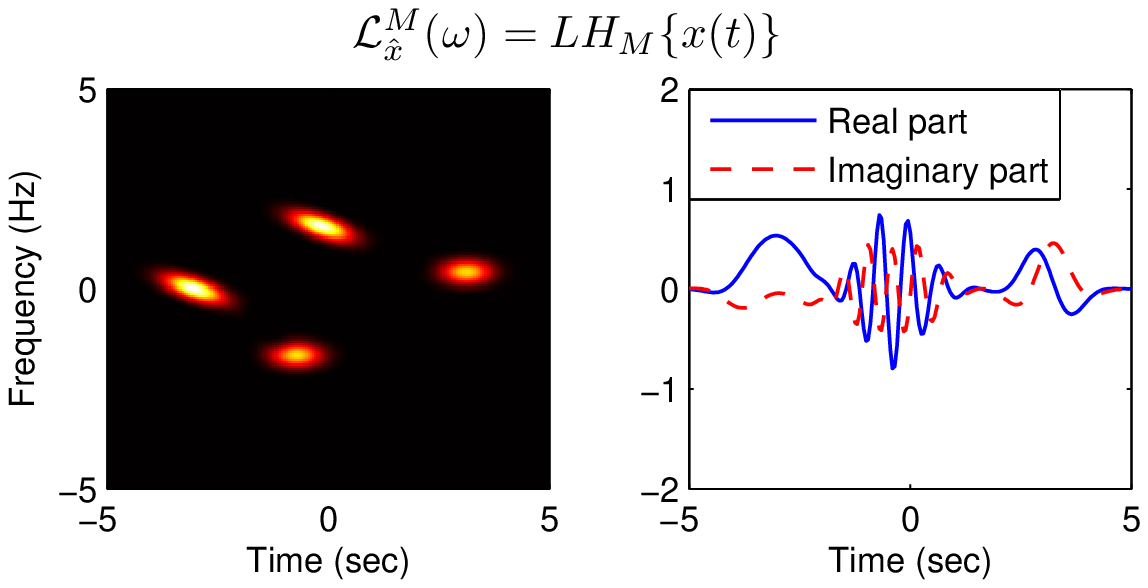}}
\\\vspace*{3pt}
\subfigure[] {
\centering
\includegraphics[width=0.95\columnwidth]{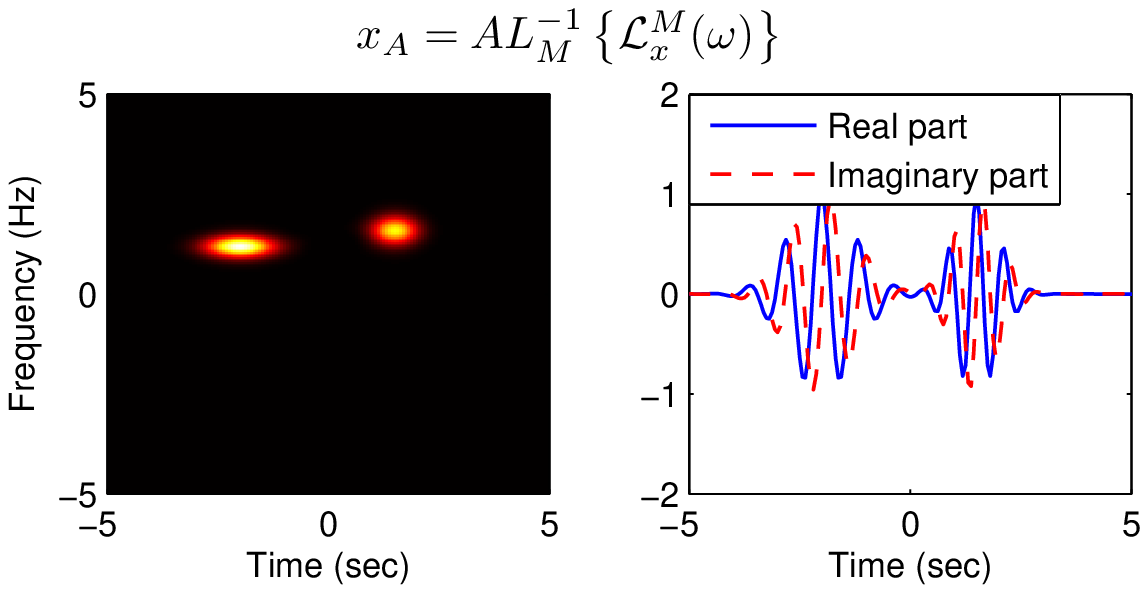}}
\quad
\subfigure[] {
\centering
\includegraphics[width=0.95\columnwidth]{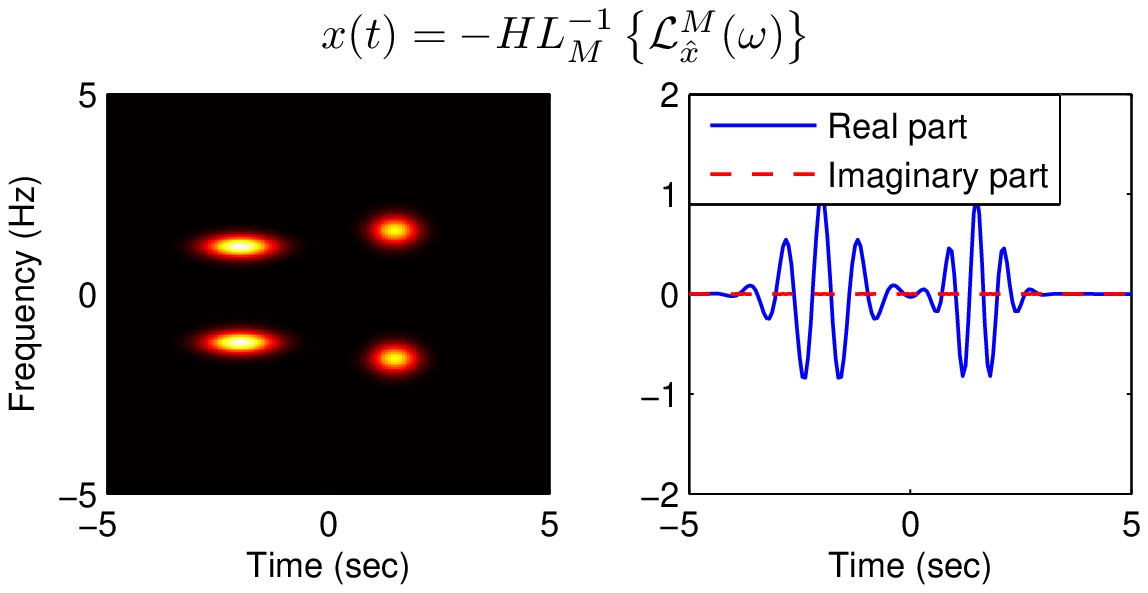}}
\\\vspace*{3pt}
\subfigure[] {
\centering
\includegraphics[width=0.95\columnwidth]{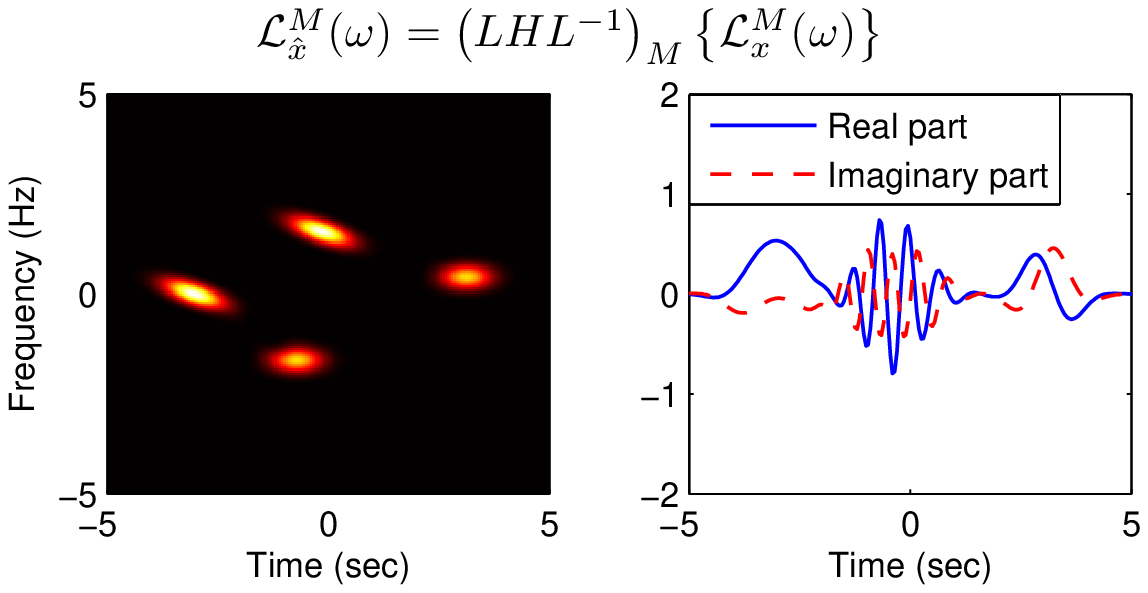}}
\quad
\subfigure[] {
\centering
\includegraphics[width=0.95\columnwidth]{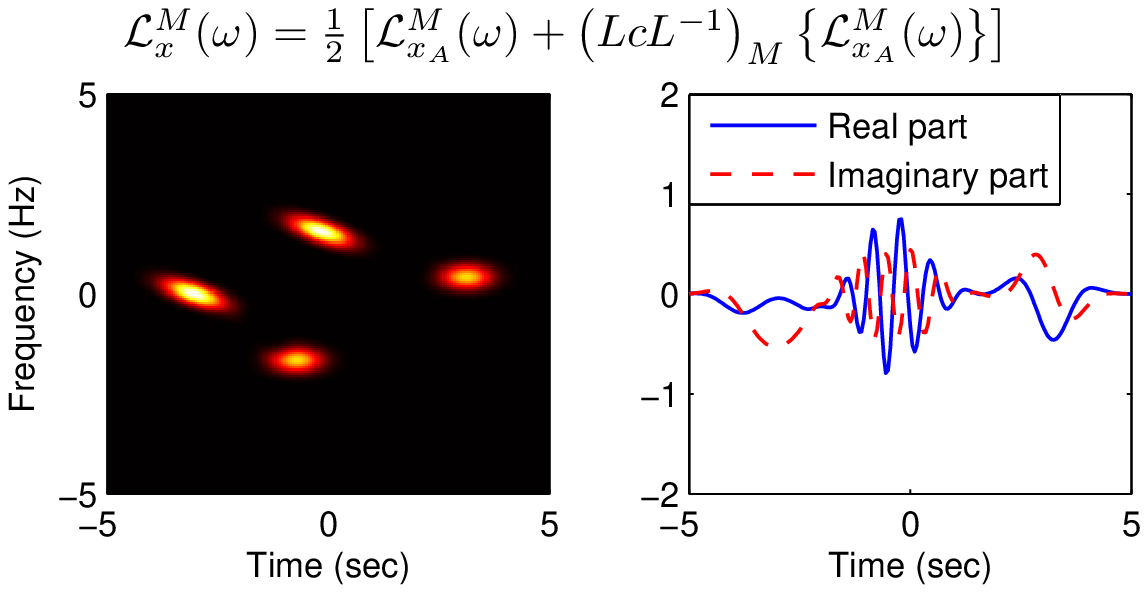}}
}
\vspace*{-0.25cm}
\caption{For $a\neq0$ ($M=(a,b,c,d)=(0.8,1.2,-0.4,0.65)$), the signals generated by the 
joint transforms:
(a) ${\cal L}^M_{x_A}(\omega)$ generated by $jL{A_M}\left\{ {\hat x}(t) \right\}$;
(b) ${\cal L}^M_{\hat x}(\omega)$ generated by  $L{H_M}\{x(t)\}$;
(c) $x_A(t)$ generated by $AL_M^{ - 1}\left\{ {{\cal L}^M_x}(\omega) \right\}$;
(d) $x(t)$ generated by $- HL_M^{ - 1}\left\{ {{\cal L}_{\hat x}^M(\omega)} \right\}$;
(e) ${{\cal L}^M_{\hat x}}(\omega)$ generated by $\left(LHL^{-1}\right)_M\left\{ {{\cal L}^M_x}(\omega) \right\}$;
(f) ${\cal L}^M_x(\omega)$ generated by $\frac{1}{2}\left[ {{\cal L}^M_{x_A}(\omega) + \left(LcL^{-1}\right)_M\left\{ {{\cal L}^M_{x_A}(\omega)} \right\}} \right]$. The \textbf{left} subfigures in (a)-(f) show the 
time-frequency distributions, while the \textbf{right} subfigures in (a)-(f) show the
real and imaginary parts of the time waveforms.
}
\label{fig:Sim2}
\vspace*{-0.4cm}\end{figure*}

\mysubsection{Joint LCT-Conjugation-ILCT ($b\neq0$)}\label{subsec:LcI}
Since the transformations from ${\cal L}^M_{x_A}(\omega)$ to ${\cal L}^M_{x}(\omega)$ and
from ${\cal L}^M_{x_A}(\omega)$ to
${\cal L}^M_{\hat x}(\omega)$ can be respectively realized by ${\cal L}^M_{x}(\omega)=\frac{1}{2}\left({\cal L}^M_{x_A}(\omega)+{\cal L}^M_{x_A^*}(\omega)\right)$ and ${\cal L}^M_{\hat x}(\omega)=\frac{1}{2j}\left({\cal L}^M_{x_A}(\omega)-{\cal L}^M_{x_A^*}(\omega)\right)$, the joint LCT-conjugation-ILCT is introduced.
The joint LCT-conjugation-ILCT, denoted by $\left(LcL^{-1}\right)_M$, is an operator such that ${\cal L}^M_{x_A}(\omega)$ can be transformed into ${\cal L}^M_{x_A^*}(\omega)$ and vice versa. 
If the input is ${\cal L}^M_{x_A}(\omega)$, it is equivalent to transforming ${\cal L}^M_{x_A}(\omega)$ into $x_A(t)$ by the ILCT first, then calculating the complex conjugate of $x_A(t)$ (i.e., $x^*_A(t)$), and finally transforming $x^*_A(t)$ into ${\cal L}^M_{x_A^*}(\omega)$ by the LCT,
\begin{align}\label{eq:LcI00}
{\cal L}^M_{x_A^*}(\omega)&=\left(LcL^{-1}\right)_M\left\{ {\cal L}^M_{x_A}(\omega) \right\} \nn\\ &\triangleq{{\cal L}^M}\left\{ {{{\left[ {{{\cal L}^{{M^{ - 1}}}}\left\{ {{\cal L}_{x_A}^M(\omega )} \right\}} \right]}^{\rm{*}}}} \right\}.
\end{align}
According to the conjugate property of the LCT in (\ref{eq:LCT08}), formula (\ref{eq:LcI00}) is equivalent to
\begin{equation}\label{eq:LcI02}
\left[{\cal L}^{(a, - b, - c,d)}\left\{ {{\cal L}^{(d, - b, - c,a)}\left\{ {{\cal L}_{x_A}^M(\omega )} \right\}} \right\}\right]^*.
\end{equation}
Applying the LCT Form I in (\ref{eq:LCT02}) to (\ref{eq:LcI02}) yields (\ref{eq:LcI04}).
\setcounter{equation}{\value{count2}}
If $a=0$, (\ref{eq:LcI04}) can be simplified to
\begin{align}\label{eq:LcI06}
\left(LcL^{-1}\right)_M\left\{ {{\cal L}_{x_A}^M(\omega )} \right\}
= - j{\mathop{\rm sgn}} (b)\ {e^{j\pi \frac{{2d}}{b}{\omega ^2}}}{\left[ {{\cal L}_{x_A}^M( - \omega )} \right]^*}.
\end{align}
If $a\neq0$, substituting (\ref{eq:Int04}) with $\gamma=-\frac{2a}{b}$ and $f=-\frac{\omega+\eta}{b}$ into (\ref{eq:LcI04}) yields
\begin{align}\label{eq:LcI08}
&\left(LcL^{-1}\right)_M\left\{ {{\cal L}_{x_A}^M(\omega )} \right\}\nn\\
&=\frac{1}{{jb}}\sqrt {\frac{b}{{ - j2a}}} {e^{j\pi \frac{{2ad - 1}}{{2ab}}{\omega ^2}}}\hspace{-5pt}\int\limits_{ - \infty }^\infty  {{{\left[ {{\cal L}_{x_A}^M(\eta )} \right]}^*}{e^{j\pi \frac{{2ad - 1}}{{2ab}}{\eta ^2}}}} {e^{-j\pi \frac{\omega }{{ab}}\eta }}d\eta.
\end{align}

\vspace*{-0.5cm}
\mysubsection{Summary}\label{subsec:Sum}
The joint LCT-AS, joint LCT-HT, joint AS-ILCT, joint HT-ILCT, joint LCT-HT-ILCT and joint LCT-conjugation-ILCT are 
given respectively in (\ref{eq:LA01}), (\ref{eq:LH06}), (\ref{eq:AI02}), (\ref{eq:HI04}), (\ref{eq:LHI10}) and (\ref{eq:LcI06}) for $a=0$; and
respectively in (\ref{eq:LA08}), (\ref{eq:LH06}), (\ref{eq:AI12}), (\ref{eq:HI08}), (\ref{eq:LHI10}) and (\ref{eq:LcI08}) for $a\neq0$.
It can be found that the joint transforms involve only one integral or even no integral.
Besides, for $a\neq0$, four of the six joint transforms are carried out by $g_1(t)$ and/or $g_2(t)$, which are pre-computable.
Therefore, the joint transforms can reduce computational complexity of the relationships involving multiple integral transforms, which are shown in Fig.~\ref{fig:JT23}.
The symbol $I$ denotes the identity operator.
For completeness, a joint transform related to the LCT of $x_A^*(t)$, called joint LCT-conjugation-AS, is also proposed.
Interested readers can refer to Appendix \ref{App:LcA}.

\begin{figure*}[t]
\addtolength{\subfigcapskip}{-7pt}
\centering{
\subfigure[] {
\centering
\includegraphics[width=0.95\columnwidth]{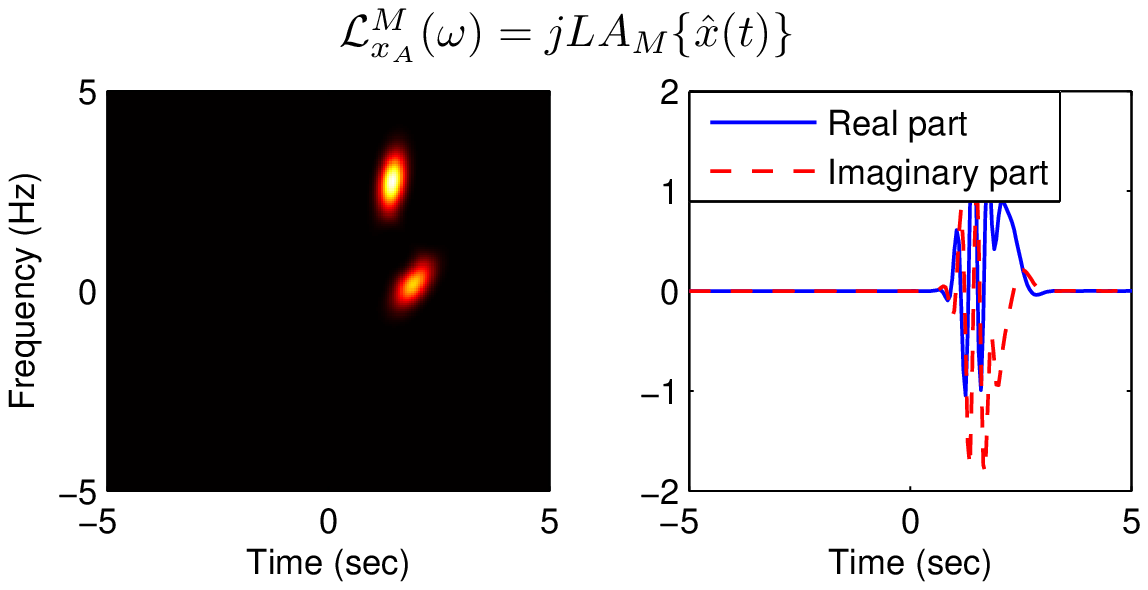}}
\quad
\subfigure[] {
\centering
\includegraphics[width=0.95\columnwidth]{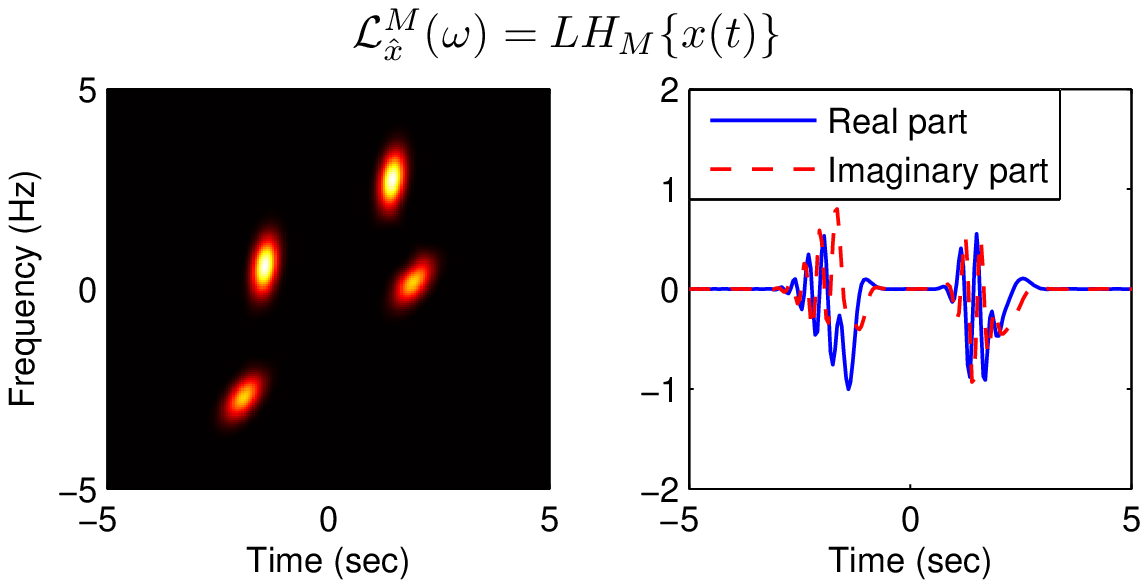}}
\\\vspace*{3pt}
\subfigure[] {
\centering
\includegraphics[width=0.95\columnwidth]{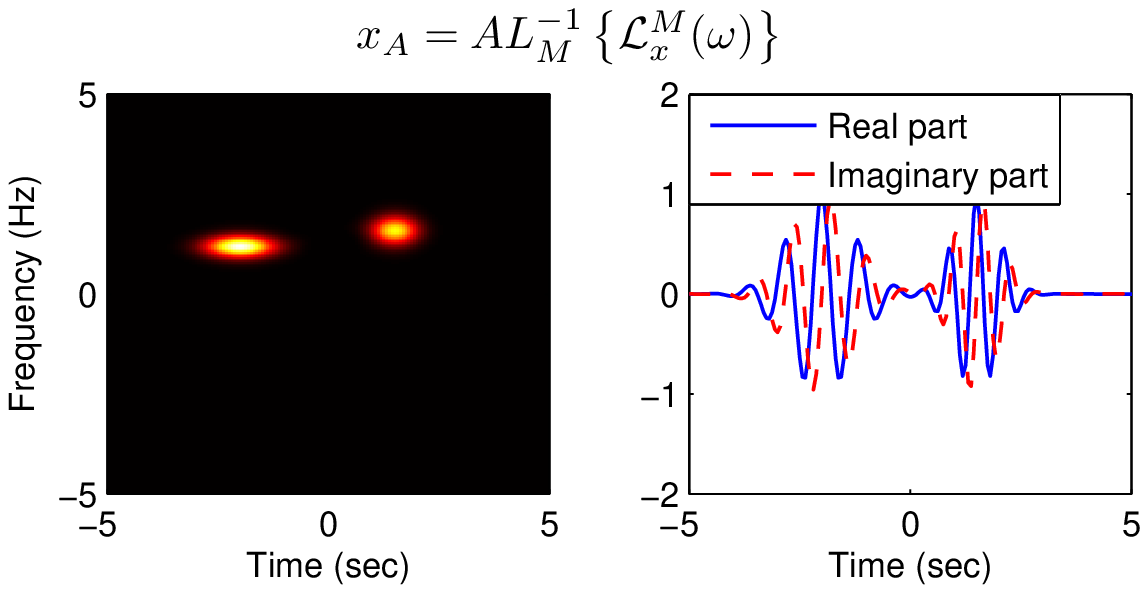}}
\quad
\subfigure[] {
\centering
\includegraphics[width=0.95\columnwidth]{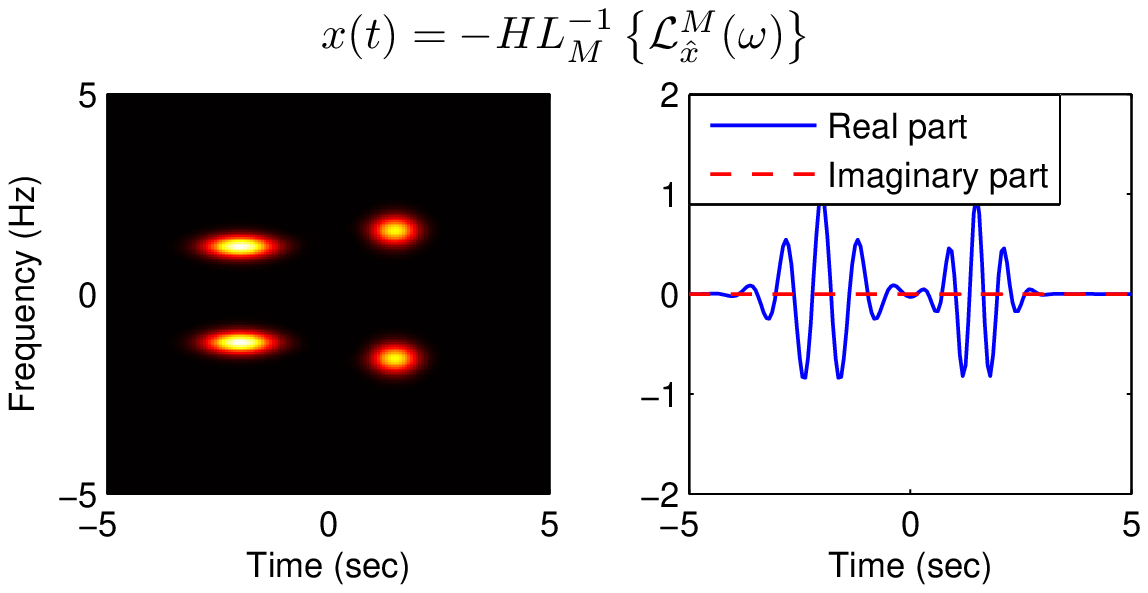}}
\\\vspace*{3pt}
\subfigure[] {
\centering
\includegraphics[width=0.95\columnwidth]{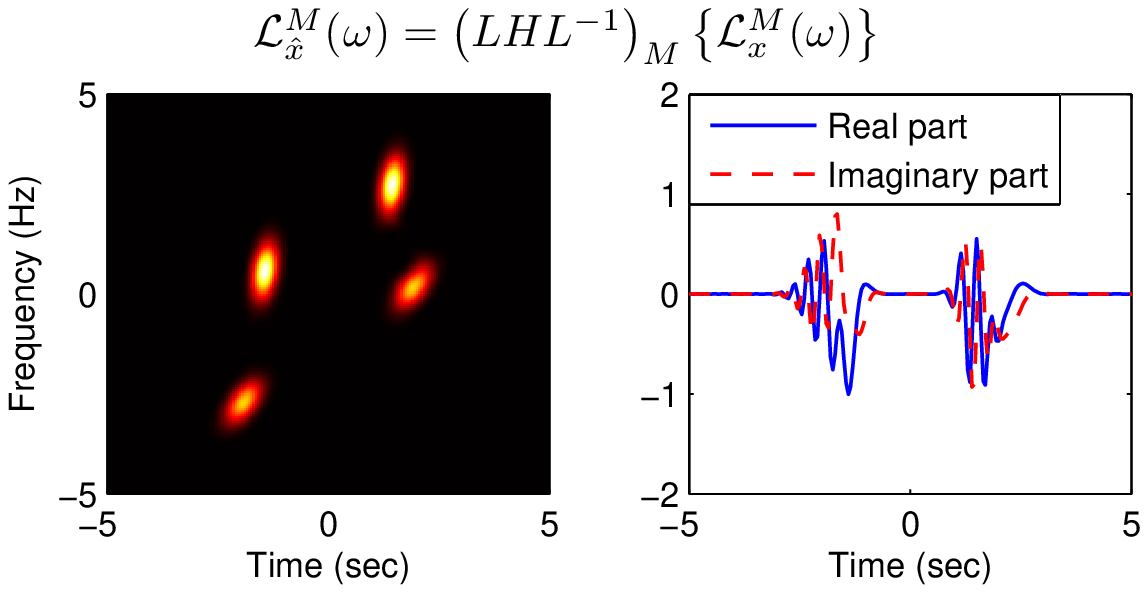}}
\quad
\subfigure[] {
\centering
\includegraphics[width=0.95\columnwidth]{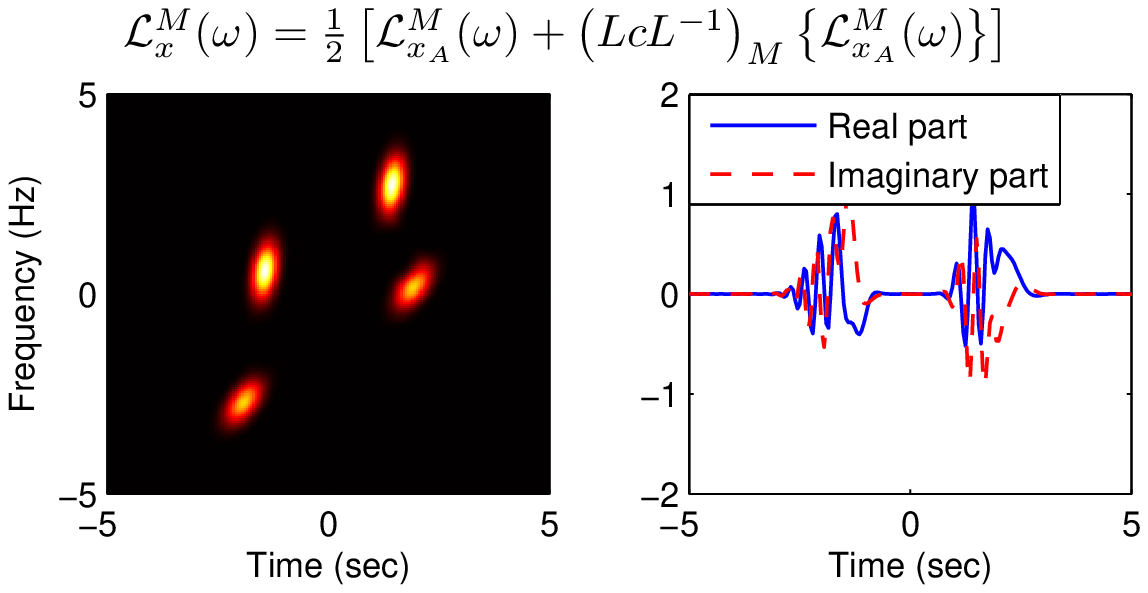}}
}
\caption{For $a=0$ 
($M=(a,b,c,d)=(0,1.2,0.833,0.9)$), the signals generated by the 
joint transforms:
(a) ${\cal L}^M_{x_A}(\omega)$ generated by $jL{A_M}\left\{ {\hat x}(t) \right\}$;
(b) ${\cal L}^M_{\hat x}(\omega)$ generated by  $L{H_M}\{x(t)\}$;
(c) $x_A(t)$ generated by $AL_M^{ - 1}\left\{ {{\cal L}^M_x}(\omega) \right\}$;
(d) $x(t)$ generated by $- HL_M^{ - 1}\left\{ {{\cal L}_{\hat x}^M(\omega)} \right\}$;
(e) ${{\cal L}^M_{\hat x}}(\omega)$ generated by $\left(LHL^{-1}\right)_M\left\{ {{\cal L}^M_x}(\omega) \right\}$;
(f) ${\cal L}^M_x(\omega)$ generated by $\frac{1}{2}\left[ {{\cal L}^M_{x_A}(\omega) + \left(LcL^{-1}\right)_M\left\{ {{\cal L}^M_{x_A}(\omega)} \right\}} \right]$. The \textbf{left} subfigures in (a)-(f) show the 
time-frequency distributions, while the \textbf{right} subfigures in (a)-(f) show the
real and imaginary parts of the time waveforms.
}
\label{fig:Sim4}
\end{figure*}

\mysection{Simulation Results}\label{sec:Sim}
First, a simulation is given to illustrate why the proposed 
joint transform is more preferred than the PHT \cite{fu2008} in practice.
Then, the six proposed joint transforms are verified by numerical simulations.

\mysubsection{Proposed Joint LCT-AS Compared With the PHT}\label{subsec:SimP}
Consider a baseband real signal $x(t)$ consisting of a sinc function and a modulated Gaussian function,
\begin{align}\label{eq:Sim00}
x(t) = \frac{11}{5}\mathrm{sinc}\left(\frac{11}{2}\left(t+\frac{3}{2}\right)\right)
+&{e^{ -2{{(t - 2)}^2}}} \cos ( {2\pi t}).
\end{align}
The time-frequency distribution (TFD) of $x(t)$
is depicted in Fig.~\ref{fig:PHT}(a).
Here, the quadratic TFD called deconvolutive short-time 
Fourier
transform (DSTFT) \cite{lu2009} is adopted, which has almost no cross-terms and can be easily implemented by MATLAB.
The TFD of the LCT of $x(t)$ (i.e., ${{\cal L}^M_x}(\omega)$) is given in Fig.~\ref{fig:PHT}(b).
In this simulation, $M=(a,b,c,d)=(0.8,1.2,-0.4,0.65)$ is adopted for LCT.
The LCT of the generalized analytic signal based on PHT is equivalent to suppressing the negative part of ${{\cal L}^M_x}(\omega)$, as shown in Fig.~\ref{fig:PHT}(c).
It is impossible to recover $x(t)$ from the signal in Fig.~\ref{fig:PHT}(c) without distortion because of the significant information loss of the sinc component.
In contrast, the signal generated by the proposed joint LCT-AS contains the whole information of $x(t)$.
The TFD of the conventional analytic signal of $x(t)$ (i.e., $x_A(t)$) is depicted in Fig.~\ref{fig:PHT}(d).
The LCT of $x_A(t)$ (i.e., ${{\cal L}^M_{x_A}}(\omega)$) can be calculated directly from $x(t)$ by the joint LCT-AS, as shown in Fig.~\ref{fig:PHT}(e).
It is obvious that the joint LCT-AS is reversible and undistorted.

\vspace*{-0.2cm}
\mysubsection{Verifying the Proposed Joint Transforms Numerically}\label{subsec:SimV}
According to the derivations of the joint transforms in sections from \ref{subsec:LA} to \ref{subsec:LcI}, it is obvious that each joint transform is mathematically equivalent to the corresponding cascade of multiple integral transforms.
In the following, we once again verify these joint transforms by numerically simulations.
There are six kinds of joint transforms.
Therefore, six of the 14 relationships depicted in Fig.~\ref{fig:JT23} are chosen:
${\hat x}(t)\longrightarrow{\cal L}^M_{x_A}(\omega)$,
$x(t)\longrightarrow{\cal L}^M_{\hat x}(\omega)$,
${{\cal L}^M_x}(\omega)\longrightarrow x_A(t)$,
${{\cal L}_{\hat x}^M(\omega)}\longrightarrow x(t)$,
${{\cal L}^M_x}(\omega)\longrightarrow{{\cal L}^M_{\hat x}}(\omega)$ and
${{\cal L}^M_{x_A}(\omega)} \longrightarrow{\cal L}^M_x(\omega)$.
For both cases $a\neq0$ and $a=0$, 
numerical simulations are given to check whether the following six equalities are true:
\begin{align}\label{eq:Sim02}
\rm{(a):}&\ \  {\cal L}^M\left\{j{\cal A}\{{\hat x}(t)\}\right\}=jL{A_M}\left\{ {\hat x}(t) \right\} \nn\\
\rm{(b):}&\ \  {{\cal L}^M}\left\{{\cal H}\{x(t)\}\right\}=L{H_M}\{x(t)\}\nn\\
\rm{(c):}&\ \  {\cal A}\left\{ {{{\cal L}^{M^{-1}}}\left\{ {{\cal L}^M_x}(\omega) \right\}} \right\}=AL_M^{ - 1}\left\{ {{\cal L}^M_x}(\omega) \right\} \nn\\
\rm{(d):}&\ \  - {\cal H}\left\{ {{{\cal L}^{ M^{- 1}}}\left\{ {{\cal L}_{\hat x}^M(\omega)} \right\}} \right\}=- HL_M^{ - 1}\left\{ {{\cal L}_{\hat x}^M(\omega)} \right\} \\
\rm{(e):}&\ \  {{\cal L}}\left\{ {{\cal H}\left\{ {{{\cal L}^{{-1}}}\left\{ {{\cal L}^M_x}(\omega) \right\}} \right\}} \right\}=\left(LHL^{-1}\right)_M\left\{ {{\cal L}^M_x}(\omega) \right\} \nn\\
\rm{(f):}&\ \  \frac{1}{2}\left[ {\cal L}^M_{x_A}(\omega) + {{\cal L}^M}\left\{ {{{\left[ {{{\cal L}^{{M^{ - 1}}}}\left\{ {{\cal L}_{x_A}^M(\omega )} \right\}} \right]}^{\rm{*}}}} \right\}\right]\nn\\
&\quad\qquad\qquad=\frac{1}{2}\left[ {{\cal L}^M_{x_A}(\omega) + \left(LcL^{-1}\right)_M\left\{ {{\cal L}^M_{x_A}(\omega)} \right\}} \right].\nn
\end{align}

Consider a real signal consisting of two modulated Gaussian signals 
with different variances and different carrier frequencies,
\begin{align}\label{eq:Sim04}
x(t) = &{e^{ - \pi \frac{{13}}{{45}}{{(t + 2)}^2}}} \cos \left( {2\pi \frac{6}{5}(t + 2)} \right) \nn\\
&\quad\qquad+ {e^{ - \pi \frac{{16}}{{25}}{{\left( {t - \frac{3}{2}} \right)}^2}}}\cos \left( {2\pi \frac{8}{5}\left( {t - \frac{3}{2}} \right)} \right).
\end{align}
For $a\neq0$, $M=(a,b,c,d)=(0.8,1.2,-0.4,0.65)$ is adopted.
First, ${\hat x}(t)$, $x_A(t)$, ${{\cal L}^M_x}(\omega)$, ${{\cal L}^M_{\hat x}}(\omega)$ and ${{\cal L}^M_{x_A}}(\omega)$ are constructed from $x(t)$ by 
the HT, analytic signal and LCT (see Fig.~\ref{fig:JT1}).
These 
six signals are then used as the inputs of (\ref{eq:Sim02}).
The 
outputs of the right-hand side transformations of (\ref{eq:Sim02}) (i.e., 
the joint transforms) are 
illustrated in Fig.~\ref{fig:Sim2}(a) to (f),
respectively.
The left subfigures in (a)-(f) show the TFDs, while the right subfigures in (a)-(f) show the
real and imaginary parts of the time waveforms.
The outputs of the left-hand side transformations of (\ref{eq:Sim02}) (i.e., the cascades of integral transforms) are not shown here, because the maximal differences 
with the 
outputs of the joint transforms are all smaller than $10^{-7}$, which may be caused by
numerical round-off error.
Therefore, it can be concluded that the six equalities in (\ref{eq:Sim02}) are true.

The six kinds of joint transforms have been 
numerically verified for $a\neq0$.
Next, the case that $a=0$ is concerned.
Again, the signal $x(t)$ in (\ref{eq:Sim04}) is used, and the same six relationships are chosen.
Except that 
$M=(a,b,c,d)=(0,1.2,0.833,0.9)$, repeating the procedures mentioned in the previous paragraph yields the simulation 
result, Fig.~\ref{fig:Sim4}.
In this simulation, the maximal differences between the 
outputs of the joint transforms and the ones of the cascades of integral transforms
are all smaller than 
$10^{-8}$.
Therefore, we can conclude that the 
derivations of the six joint transforms are also correct for $a=0$.


\mysection{Applications}\label{sec:Apc}
Since the proposed joint transforms are associated with the analytic signal and the LCT, 
several signal processing applications of the analytic signal and the LCT can be extended 
using the proposed transforms.
The advantages and flexibility of using the joint LCT-AS over using the analytic signal or using the LCT are also discussed.

\begin{figure}[t]
\centering
\includegraphics[width=0.95\columnwidth]{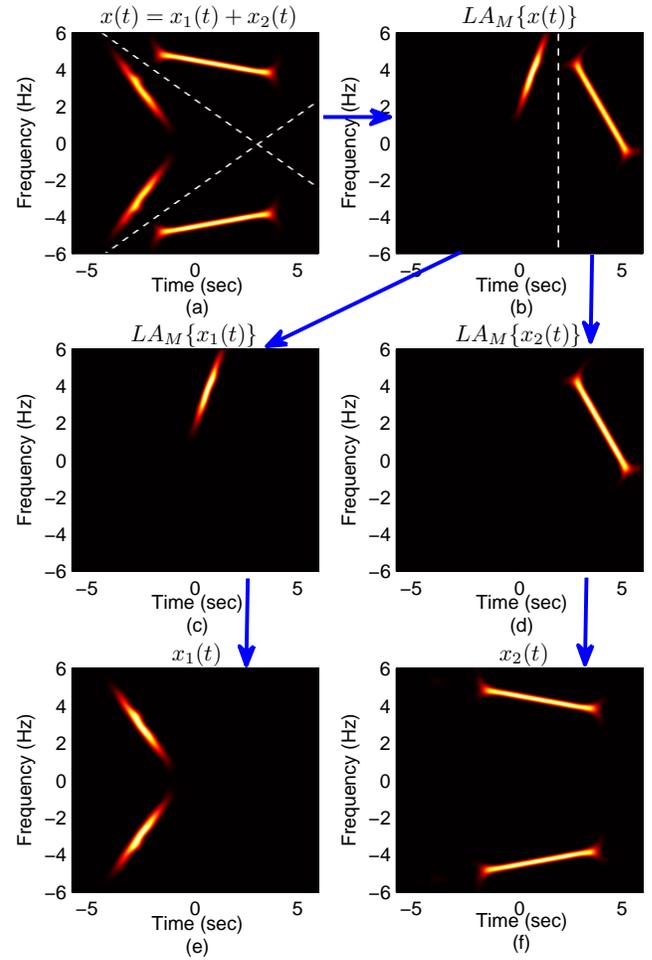}
\vspace*{-0.2cm}
\caption{
Time-frequency distributions of (a) a real signal $x(t)=x_1(t)+x_2(t)$; (b) the joint LCT-AS of $x(t)$, i.e., $LA_M\{x(t)\}$; (c) $LA_M\{x_1(t)\}$ separated out from $LA_M\{x(t)\}$; (d) $LA_M\{x_2(t)\}$ separated out from $LA_M\{x(t)\}$; (e) $x_1(t)$ recovered from the real part of the ILCT of $LA_M\{x_1(t)\}$; and (f) $x_2(t)$ recovered from the real part of the ILCT of $LA_M\{x_2(t)\}$.
}
\label{fig:Sep}
\vspace*{-0.3cm}\end{figure}

\mysubsection{Signal Separation in the LCT Domain}\label{subsec:Sep}
Consider a signal with components overlapped in the time domain.
In order to separate these components, a well-known approach is to work in the LCT (or FRFT) domain.
For example, given $x(t)=x_1(t)+x_2(t)$,
if $x_1(t)$ and $x_2(t)$ are not overlapped in the $M$-LCT domain for some parameter matrix $M$,
then they can be easily separated by ${\cal L}^M_{x}(\omega)u(\omega-\omega_0)$ and ${\cal L}^M_{x}(\omega)u(-\omega+\omega_0)$.
Now, consider $x(t)=x_1(t)+x_2(t)$ is a real signal with TFD as shown in Fig.~\ref{fig:Sep}(a).
Since two cutoff lines are required to separate $x_1(t)$ and $x_2(t)$ (see the dashed lines in Fig.~\ref{fig:Sep}(a)),
the LCT should be employed twice for two separation procedures.
Fortunately, the non-negative frequencies 
contain the whole information of $x(t)$.
If the separation is performed on the analytic signal of $x(t)$, the cutoff lines used in the negative frequency domain can be ignored.
Accordingly, in this example, one joint LCT-AS is enough for separation, as shown in Fig.~\ref{fig:Sep}(b), (c) and (d).
Also, Fig.~\ref{fig:Sep}(e) and (f) show that $x_1(t)$ and $x_2(t)$ can be recovered from $LA_M\{x_1(t)\}$ and $LA_M\{x_2(t)\}$ by the inverse transform of the joint LCT-AS (i.e., the real part of the ILCT), respectively.
For more complex signals or signals with more components, there would be more than one cutoff line in the positive frequency domain.
In such situation, besides the joint LCT-AS, additional LCTs are required for additional separation procedures.

Simply using the analytic signal is unable to separate multiple components which are overlapped in the time domain.
Combining the analytic signal and the LCT (i.e., the joint LCT-AS) can substantially reduce the number of separation procedures required by simply using the LCT.

The cutoff lines mentioned above can be determined by 
separating components on the TFD.
The parameters of $M$ can be determined by the slope of the cutoff line and the distance between the cutoff line and the origin.
One can refer to \cite{pei2007,pei2001} for more details.

\mysubsection{Filter Design in the LCT Domain}\label{subsec:Fil}
The notion of filter design by the LCT is similar to the signal separation in the LCT domain.
Consider a noisy signal $y(t)=x(t)+n(t)$, where $x(t)$ is the desired signal and $n(t)$ is noise/interference.
To filter out $x(t)$, a filter $H$ operating in the LCT domain results in
\begin{align}\label{eq:Fil02}
x(t)={\cal L}^{M^{-1}}\left\{ H(\omega)\cdot {\cal L}^M\{y(t)\}\right\}
\end{align}
where the parameter matrix $M$ can be determined by the method mentioned in the previous subsection.
If $x(t)$ is real, we can first calculate the analytic signal of $y(t)$ (i.e., $y_A(t)$) to eliminate the noise/interference components in the negative frequency domain,
and (\ref{eq:Fil02}) becomes
\begin{align}\label{eq:Fil06}
x(t)=
Re\left\{{\cal L}^{M^{-1}}\left\{ H(\omega)\cdot LA_M\{y(t)\}\right\}\right\}.
\end{align}
Since the number of noise/interference components is reduced, the design of the filter $H$  and the parameter matrix $M$ becomes more flexible.

This application again shows that combining the analytic signal and the LCT (i.e., joint LCT-AS) can benefit from both the advantage
of the analytic signal (i.e., eliminating the noise/interference components in the negative frequency domain) and
the flexibility of the LCT (i.e., filter design in the LCT domain).

\begin{figure}[t]
\centering
\includegraphics[width=\columnwidth]{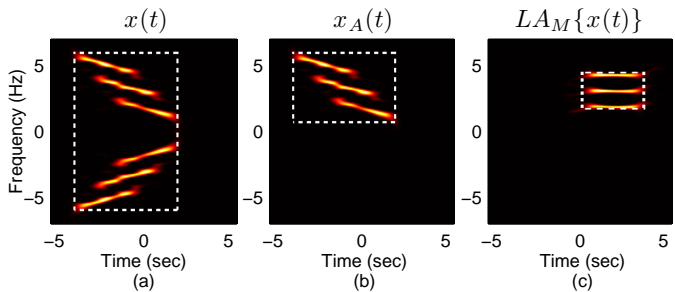}
\vspace*{-0.2cm}
\caption{
Time-frequency distributions of (a) a real signal $x(t)$; (b) the analytic signal of $x(t)$; and (c) the joint LCT-AS of $x(t)$.
The areas of the dashed rectangles in (a)-(c) represent the bandwidth-time products, denoted by $B_aT_a$, $B_bT_b$, and $B_cT_c$, respectively.
It can be found that $B_aT_a>2B_bT_b>2B_cT_c$.
}
\label{fig:Sam}
\vspace*{-0.3cm}
\end{figure}

\mysubsection{Sampling Using the Joint Transforms}\label{subsec:Sam}
Consider a signal $z(t)$ which is time-limited to $[t_0,t_1]$ and approximately band-limited to $[f_0,f_1]$.
In order to reduce the number of sampling samples, $z(t)$ is  frequency shifted to baseband $[-(f_1-f_0)/2,(f_1-f_0)/2]$, which yields that the minimal sampling rate for perfect reconstruction is $f_s=f_1-f_0$.
It follows that the minimal number of samples is $(t_1-t_0)(f_1-f_0)=BT$, where $BT$ denotes the bandwidth-time (BT) product.
If the baseband signal is real, the number of \emph{real} samples is $N=BT$; otherwise, $N=2BT$.

Given a real signal $x(t)$ with TFD as shown in Fig.~\ref{fig:Sam}(a), 
the BT product of $x(t)$, denoted by $B_aT_a$,  is equal to the area of the dashed rectangle in Fig.~\ref{fig:Sam}(a).
Since $x(t)$ is real, the number of real samples is also $B_aT_a$.
To reduce the number of samples, a simple method is to sample the analytic signal of $x(t)$, i.e., $x_A(t)$.
The TFD of $x_A(t)$ is depicted in Fig.~\ref{fig:Sam}(b), and the corresponding BT product $B_bT_b$ is the area of the dashed rectangle in Fig.~\ref{fig:Sam}(b).
Since the baseband signal of $x_A(t)$ is complex, the number of real samples is $2B_bT_b$.
We can find out that $2B_bT_b<B_aT_a$.

With appropriate parameter matrix $M$, the LCT can further reduce the number of samples of $x_A(t)$.
In order to find the 
appropriate $M$, we first search the optimal shear factor $\alpha$ such that the horizontal shearing with $\alpha$ on the TFD of $x_A(t)$ yields the minimal time duration.
After horizontal shearing, find the optimal shear factor $\beta$ such that the vertical shearing with $\beta$ leads to the minimal bandwidth.
Then, the 
appropriate $M$ for small BT product is given by
\begin{align}\label{eq:Sam02}
M=\begin{bmatrix}
1\ &0\\
\beta\ &1
\end{bmatrix}\begin{bmatrix}
1\ &\alpha\\
0\ &1
\end{bmatrix}.
\end{align}
The TFD of the $M$-LCT of $x_A(t)$, i.e., the joint LCT-AS of $x(t)$, is depicted in Fig.~\ref{fig:Sam}(c).
The area of the dashed rectangle in Fig.~\ref{fig:Sam}(c) is denoted by $B_cT_c$, and it follows that the number of real samples is $2B_cT_c$.
It can be found that $2B_cT_c<2B_bT_b$.
Therefore, when sampling a real signal, the joint LCT-AS outperforms the analytic signal in reducing the number of samples.

If we want to sample the LCT of a real signal, i.e., ${\cal L}^M_{x}(\omega)$, we can transform ${\cal L}^M_{x}(\omega)$ into $x_A(t)$ by $AL_M^{-1}$ or into ${\cal L}^M_{x_A}(\omega)$ by $I+j(LHL^{-1})_M$ to reduce the BT product.

\mysubsection{Instantaneous Frequency and Amplitude of Joint LCT-AS}\label{subsec:IFE}
Given a monocomponent real signal $x(t)$ with the analytic signal of the form $x_A(t)=a(t)e^{j\phi(t)}$, where $a(t)\geq0$,
the instantaneous frequency (IF) of $x(t)$ is given by $f_x(t)=\frac{1}{2\pi}\phi'(t)$ \cite{boashash2003}.
For any time instant $t_i$, assume 
\begin{align}\label{eq:IFE00}
f_x(t_i)=f_i.
\end{align}
In the following, we will analyze the IF of the joint LCT-AS of $x(t)$, i.e., the LCT of $x_A(t)$.
For $b\neq0$, the $M$ in the LCT can be decomposed into four matrices:
\begin{align}\label{eq:IFE02}
M=\begin{bmatrix}
a\ &b\\
c\ &d
\end{bmatrix}=\begin{bmatrix}
1\ &0\\
d/b\ &1
\end{bmatrix}\begin{bmatrix}
0\ &1\\
-1\ &0
\end{bmatrix}\begin{bmatrix}
1\ &0\\
ab\ &1
\end{bmatrix}\begin{bmatrix}
1/b\ &0\\
0\ &b
\end{bmatrix}
\end{align}
which implies the LCT can be decomposed into four operations: scaling, chirp multiplication, Fourier transform and chirp multiplication.
\begin{itemize}
\item 
Step 1: scaling with parameter $b$,
\begin{align}\label{eq:IFE04}
y(t)=\sqrt{b}\ x_A(bt)=\sqrt{b}\ a(bt)e^{j\phi(bt)}.
\end{align}
The IF of $y(t)$, denoted by ${f_y}(t)$, is given by
\begin{align}\label{eq:IFE06}
{f_y}(t) = \frac{1}{{2\pi }}\frac{\partial }{{\partial t}}\phi (bt) = b\frac{1}{{2\pi }}\phi '(bt) = b{f_x}(bt).
\end{align}

\item
Step 2: chirp multiplication with chirp rate $ab$,
\begin{align}\label{eq:IFE08}
z(t)=e^{j\pi abt^2}y(t)=\sqrt{b}\ a(bt)e^{j\pi abt^2+j\phi(bt)}.
\end{align}
The IF of $z(t)$, denoted by ${f_z}(t)$, is given by
\begin{align}\label{eq:IFE10}
{f_z}(t) = abt +{f_y}(t) =   abt+b{f_x}(bt).
\end{align}
From (\ref{eq:IFE00}), as $t=t_i/b$, (\ref{eq:IFE10}) becomes 
\begin{align}\label{eq:IFE11}
f_z(t_i/b)=at_i+bf_x(t_i)=at_i+bf_i.
\end{align}

\item
Step 3: Fourier transform,
\begin{align}\label{eq:IFE12}
Z(\omega)=\sqrt{-j}\cdot{\cal F}\{z(t)\}.
\end{align}
As $\sqrt{-j}=e^{-j\pi/4}$, assume $Z(\omega)=A(\omega)e^{j\psi(\omega)-j\pi/4}$, where $A(\omega)\geq0$.
The IF of $Z(\omega)$, denoted by $\nu_Z(\omega)$, is given by 
\begin{align}\label{eq:IFE13_1}
\nu_Z(\omega)=\frac{1}{2\pi}\psi'(\omega)=-\tau_z(\omega)
\end{align}
where $\tau_z(\omega)$ is known as the \emph{group delay} (GD) of $z(t)$.
If the BT of $z(t)$ is large and the IF of $z(t)$ is monotonic, the GD is the inverse function of the IF, i.e., $\tau_z=f^{-1}_z$ \cite{boashash921}.
Therefore, 
\begin{align}\label{eq:IFE13_1}
\nu_Z(\omega)=-f^{-1}_z(\omega)
\end{align}
and from (\ref{eq:IFE11}), it follows that 
\begin{align}\label{eq:IFE14}
\nu_Z(at_i+bf_i)=-f^{-1}_z(at_i+bf_i)=-t_i/b.
\end{align}

\item
Step 4: Chirp multiplication with chirp rate $d/b$,
\begin{align}\label{eq:IFE16}
{\cal L}^M_{x_A}(\omega)=e^{j\pi \frac{d}{b}\omega^2}Z(\omega).
\end{align}
The IF of ${\cal L}^M_{x_A}(\omega)$, denoted by $\nu_L(\omega)$, is given by
\begin{align}\label{eq:IFE18}
\nu_L(\omega) = \frac{d}{b}\omega +{\nu_Z}(\omega).
\end{align}
From (\ref{eq:IFE14}), for $\omega=at_i+bf_i$, we have 
\begin{align}\label{eq:IFE20}
\nu_L(at_i+bf_i)&=\frac{d}{b}(at_i+bf_i) +{\nu_Z}(at_i+bf_i) \nn\\
&=\frac{ad}{b}t_i+df_i-\frac{1}{b}t_i=ct_i+df_i.
\end{align}
\end{itemize}
In (\ref{eq:IFE20}), define $\omega_i=at_i+bf_i$ and $\nu_i=ct_i+df_i$.
If the IF of ${\cal L}^M_{x_A}(\omega)$ is $\nu_L(\omega_i)=\nu_i$, then the IF of $x(t)$ is $f_x(t_i)=f_i$ where
\begin{align}\label{eq:IFE22}
\begin{bmatrix}
t_i\\
f_i
\end{bmatrix}=\begin{bmatrix}
a\ &b\\
c\ &d
\end{bmatrix}^{-1}\begin{bmatrix}
\omega_i\\
\nu_i
\end{bmatrix}=\begin{bmatrix}
d\ &-b\\
-c\ &a
\end{bmatrix}\begin{bmatrix}
\omega_i\\
\nu_i
\end{bmatrix}.
\end{align}

\begin{figure}[t]
\centering
\includegraphics[width=0.95\columnwidth]{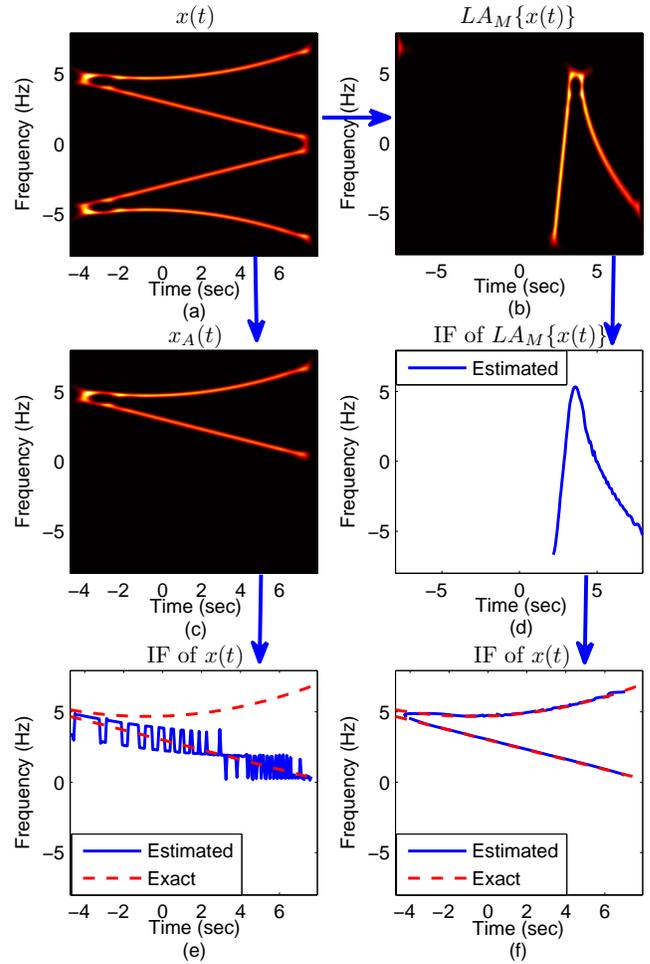}
\vspace*{-0.2cm}
\caption{
Time-frequency distributions of (a) a real signal $x(t)$; (b) the joint LCT-AS of $x(t)$, i.e., $LA_M\{x(t)\}$; and (c) the analytic signal of $x(t)$, i.e., $x_A(t)$.
(d) IF of $LA_M\{x(t)\}$. (e) IF of $x(t)$ estimated from the IF of $x_A(t)$. (f) IF of $x(t)$ estimated from the IF of $LA_M\{x(t)\}$.
}
\label{fig:IFE}
\vspace*{-0.3cm}\end{figure}
Consider a real signal $x(t)=x_1(t)+x_2(t)$ 
with $x_1(t)$ and $x_2(t)$ overlapped in the time domain.
The TFDs of $x(t)$ and its analytic signal $x_A(t)$ are depicted in Fig.~\ref{fig:IFE}(a) and (c), respectively.
Since there are two IFs at each time instant as shown in Fig.~\ref{fig:IFE}(e), the IF obtained from
the phase of $x_A(t)$ is a nonlinear combination of the two IFs, as depicted in Fig.~\ref{fig:IFE}(e).
Fortunately, as discussed above, the IF of $x(t)$ can also be estimated from
the IF of the joint LCT-AS of $x(t)$.
Therefore, the primary task is finding an appropriate $M$ for the LCT.
Observing the TFD of $x(t)$, there are four ridges at each time instant.
Clockwise rotate the TFD until there is only one ridge at each time instant.
If the final rotation angle is $\alpha$, we have $M=(\cos\alpha, \sin\alpha, -\sin\alpha ,\cos\alpha)$, and the TFD of ${\cal L}^M_{x_A}(\omega)$ is given
in Fig.~\ref{fig:IFE}(b).
Next, the IF of ${\cal L}^M_{x_A}(\omega)$ is calculated,
which is a monodic function,
as depicted in Fig.~\ref{fig:IFE}(d).
Finally, according to (\ref{eq:IFE22}), the IF of $x(t)$ can be estimated from the IF of ${\cal L}^M_{x_A}(\omega)$, as shown in Fig.~\ref{fig:IFE}(f).
Therefore, the joint LCT-AS is more powerful than the analytic signal for IF estimation.

Assume $x_A(t)=a(t)e^{j\phi(t)}$ and ${\cal L}^M_{x_A}(\omega)=A(\omega)e^{j\psi(\omega)}$, where $a(t)$ and $A(\omega)$ are the amplitudes of $x_A(t)$ and ${\cal L}^M_{x_A}(\omega)$, respectively.
When $b=0$, ${\cal L}^M_{x_A}(\omega)=\sqrt d \cdot{e^{j\pi \,cd\,{\omega ^2}}}x_A(d\omega )$, and thus $A(\omega)=|\sqrt d| \cdot a(d\omega)$.
When $b\neq0$, 
one cannot obtain $a(t)$ only from $A(\omega)$ without the phase information $\psi(\omega)$.
As mentioned in Section~\ref{sec:JT}, the LCT-AS can be deemed as the generalization of the analytic signal.
When the LCT is the identity operator, the LCT-AS is 
reduced to the analytic signal, and thus $A(\omega)=a(\omega)$.
When the LCT is the Fourier transform, $A(\omega)$ shows the energy distribution of $x_A(t)$ in the frequency domain.
Generally speaking, the amplitude of the joint LCT-AS describes the energy distribution of the analytic signal in the LCT domain.

\mysubsection{Secure Single-Sideband Modulation}\label{subsec:SSB}
It has been mentioned in the introduction that the fractional HT (FHT) is useful for secure single-sideband (SSB) modulation \cite{tseng2000,tao2008}.
The joint LCT-AS can also be used for secure SSB modulation.
Given a real message signal $x(t)$, the SSB signal is given by
\begin{align}\label{eq:SSB02}
x_{ssb}(t)=Re\left\{x_A(t)\cdot e^{j2\pi f_ct}\right\}
\end{align}
where $f_c$ is the radio carrier frequency.
One can recover $x(t)$ from $x_{ssb}(t)$ if $f_c$ is known.
The secure SSB modulation based on the joint LCT-AS is defined as
\begin{align}\label{eq:SSB04}
x^M_{ssb}(\omega)=Re\left\{LA_M\{x(t)\}\cdot e^{j2\pi f_c\omega}\right\}.
\end{align}
It is apparent that $x(t)$ cannot be recovered from $x^M_{ssb}(\omega)$ unless the parameter matrix $M$ is known in advance.
Accordingly, the three parameters $a$, $b$ and $c$ in $M$ can be used as the secrete keys. (As $a$, $b$ and $c$ are known, $d$ can be obtained from $d=\frac{1+bc}{a}$.)
Since the secure SSB modulation based on the FHT has only one secret key, i.e., the
fractional order used in the FHT, the secure SSB modulation based on the joint LCT-AS offers much higher security.

If the message signal is the LCT of a real signal, i.e., ${\cal L}^M_x(\omega)$, (\ref{eq:SSB04}) can be rewritten as
\begin{align}\label{eq:SSB06}
&x^M_{ssb}(\omega)=Re\left\{\left[{\cal L}^M_x(\omega)+j{\cal L}^M_{\hat x}(\omega)\right] e^{j2\pi f_c\omega}\right\}\nn\\
=&Re\left\{\left[{\cal L}^M_x(\omega)+j(LHL^{-1})_M\{{\cal L}^M_x(\omega)\}\right] e^{j2\pi f_c\omega}\right\}.
\end{align}
In such situation, the joint LCT-HT-ILCT is used for  secure SSB modulation.

\section{Conclusion}\label{sec:Con}
It has been 
shown that most real signals, especially for baseband real signals, cannot be recovered perfectly from their generalized analytic signals, which are generated by the PHT.
Therefore, in this paper, the conventional 
HT and analytic signal associated with the LCT 
are concerned.
The relationships between the following six kinds of signals have been examined:
an arbitrary real signal, its HT and 
analytic signal, and the LCT of these three signals.
Since some relationships involve multiple integral transforms (may be the HT, 
analytic signal, LCT or ILCT), six kinds of transforms, called joint transforms, have been proposed.
Using the joint transforms to realize these relationships can reduce computational complexity.
Most importantly, all the joint transforms are 
reversible and undistorted.
All the joint transforms have also been verified by 
numerical simulations.
These simulations show that the numerical differences between the joint transforms and the cascades of integral transforms are down to $10^{-7}$ or less, which may be caused by numerical round-off error.
Besides, it has been shown that the joint transforms are useful in
several signal processing applications.
And in these applications, using the joint transform, which combines the analytic signal and the LCT, is preferred then simply using the analytic signal 
or the LCT.

\appendices
\mysection{Four Equivalent 
Expressions of the LCT ($a\neq0 ,b\neq0$)}\label{App:AF}
For $a\neq0$ and $b\neq0$, the LCT
defined in (\ref{eq:LCT02}) can also be expressed as the following three 
equivalent expressions, called LCT Form II, LCT Form III and LCT Form IV for short. (The original definition of the LCT ($b\neq0$) as shown in (\ref{eq:LCT02}) is called LCT Form I.)
\begin{itemize}
\item LCT Form I:\\
\begin{align}\label{eq:AF16}
{{\cal L}^M_x}(\omega)=
\sqrt {\dfrac{1}{{jb}}}\ {e^{j\pi \frac{d}{b}{\omega ^2}}}
\hspace{-5pt}\int\limits_{ - \infty }^\infty  {{e^{j\pi \frac{a}{b}{t^2}}}{e^{ - j2\pi \frac{\omega }{b}t}}} x(t)\ dt.
\end{align}

\item LCT Form II:\\
From (\ref{eq:AF16}), the LCT Form II is given by
\begin{align}\label{eq:AF18}
{{\cal L}^M_x}(\omega)&= \sqrt {\frac{1}{{jb}}} {e^{j\pi \frac{d}{b}{\omega ^2}}}{e^{ - j\pi \frac{1}{{ab}}{\omega ^2}}}\hspace{-5pt}\int\limits_{ - \infty }^\infty  {{e^{j\pi \frac{a}{b}{{\left( {\frac{\omega }{a} - t} \right)}^2}}}} x(t) dt\nn\\
 &= \sqrt {\frac{1}{{jb}}} \ {e^{j\pi \frac{c}{a}{\omega ^2}}}{\left[ {x(t) * {e^{j\pi \frac{a}{b}{t^2}}}} \right]_{\ t = \frac{\omega }{a}}}.
\end{align}

\item LCT Form III:\\
Substituting 
$d\cdot\tau$ for $t$ in (\ref{eq:AF16}) leads to the LCT Form III,
\begin{align}\label{eq:AF20}
&{{\cal L}^M_x}(\omega)\nn\\
&\qquad= |d|\sqrt {\frac{1}{{jb}}} \ {e^{j\pi \frac{d}{b}{\omega ^2}}}\int\limits_{ - \infty }^\infty  {{e^{ - j2\pi \frac{d}{b}\omega \tau }}{e^{j\pi \frac{{a{d^2}}}{b}{\tau ^2}}}} x(d\tau )\
d\tau \nn\\
 &\qquad= |d|\sqrt {\frac{1}{{jb}}} \left( {x(d\omega ){e^{j\pi cd{\omega ^2}}}} \right) * {e^{j\pi \frac{d}{b}{\omega ^2}}}.
\end{align}

\item LCT Form IV:\\
The LCT can be expressed as a transformation function of $X(f)$, i.e.,
\begin{align}\label{eq:AF21}
&{{\cal L}^M_x}(\omega)\nn\\
&= \sqrt {\frac{1}{{jb}}} {e^{j\pi \frac{d}{b}{\omega ^2}}}\int\limits_{ - \infty }^\infty  {{e^{j\pi \frac{a}{b}{t^2}}}{e^{ - j2\pi \frac{\omega }{b}t}}} {\int\limits_{ - \infty }^\infty  {X(f)} {e^{j2\pi ft}}df}\ dt \nn\\
&=  \sqrt {\frac{1}{{jb}}} {e^{j\pi \frac{d}{b}{\omega ^2}}}\int\limits_{ - \infty }^\infty  {X(f) {\int\limits_{ - \infty }^\infty  {{e^{j\pi \frac{a}{b}{t^2} - j2\pi \left( {\frac{\omega }{b} - f} \right)t}}} dt} }\ df
\end{align}
which can be simplified to the LCT Form IV by (\ref{eq:Int04}) with $\gamma=\frac{a}{b}$,
\begin{align}\label{eq:AF22}
{{\cal L}^M_x}(\omega)
=\sqrt {\frac{1}{{jb}}} \sqrt {\frac{b}{{ - ja}}} {e^{j\pi \frac{c}{a}{\omega ^2}}}\hspace{-6pt}\int\limits_{ - \infty }^\infty  \hspace{-2pt}{X(f){e^{ - j\pi \frac{b}{a}{f^2}}}{e^{j2\pi \frac{\omega }{a}f}}} df.
\end{align}

\end{itemize}

\mysection{Joint LCT-conjugation-AS ($b\neq0$)}\label{App:LcA}
The joint LCT-conjugation-AS, denoted by $LcA_M$, transforms $x(t)$ into ${\cal L}^M_{x^*_A}(\omega)$.
It is equivalent to calculating the analytic signal (AS) $x_A(t)$ of $x(t)$ first, then calculating the complex conjugate of $x_A(t)$ (i.e., $x^*_A(t)$), and finally transforming $x^*_A(t)$ into ${\cal L}^M_{x_A^*}(\omega)$ by the LCT,
\begin{equation}\label{eq:LcA00}
{\cal L}^M_{x^*_A}(\omega)=Lc{A_M}\{x(t)\} \triangleq {{\cal L}^M}\left\{ \left[{{\cal A}\left\{ {x(t)} \right\}} \right]^*\right\}.
\end{equation}
Alternatively, the following relationship is used,
\begin{align}\label{eq:LcA01}
{\cal L}^M_{x^*_A}(\omega)
&= {{\cal L}^M}\left\{ {x(t)}  - j{\cal H} \{{x(t)}\} \right\}\nn\\
&= {{\cal L}^M}\{x(t)\} - jLH_M\{x(t)\}.
\end{align}
Here, (\ref{eq:LCT14}) and the LCT Form II in (\ref{eq:LCT18}) are adopted for ${{\cal L}^M}\{x(t)\}$, and $LH_M\{x(t)\}$ has been derived in (\ref{eq:LH06}). Thus,
\begin{align}\label{eq:LcA02}
&LcA_M\{x(t)\}\nn\\
&= \left\{
  \begin{array}{l l}
\sqrt {\frac{1}{{jb}}} \ {e^{j\pi \frac{c}{a}{\omega ^2}}}{\left[ {x(t) * \left( {{g_1}(t) - {g_2}(t)} \right)} \right]_{\ t = \frac{\omega }{a}}},& a \ne 0\\
{\sqrt {\frac{1}{{jb}}} \ {e^{j\pi \frac{d}{b}{\omega ^2}}}\int\limits_{ - \infty }^\infty  {x(t)\ {e^{ - j2\pi \frac{\omega }{b}t}}} dt \cdot\left[- 2u\left( -{\frac{\omega }{b}} \right)\right],}& a = 0
  \end{array} \right.
\end{align}
where $g_1(t)$ and $g_2(t)$ is defined in (\ref{eq:LA10}).
The joint LCT-conjugation-AS 
is also used in the transformation from ${\hat x}(t)$ to ${\cal L}^M_{x^*_A}(\omega)$.



\begin{thebibliography}{10}
\providecommand{\url}[1]{#1}
\csname url@samestyle\endcsname
\providecommand{\newblock}{\relax}
\providecommand{\bibinfo}[2]{#2}
\providecommand{\BIBentrySTDinterwordspacing}{\spaceskip=0pt\relax}
\providecommand{\BIBentryALTinterwordstretchfactor}{4}
\providecommand{\BIBentryALTinterwordspacing}{\spaceskip=\fontdimen2\font plus
\BIBentryALTinterwordstretchfactor\fontdimen3\font minus
  \fontdimen4\font\relax}
\providecommand{\BIBforeignlanguage}[2]{{%
\expandafter\ifx\csname l@#1\endcsname\relax
\typeout{** WARNING: IEEEtran.bst: No hyphenation pattern has been}%
\typeout{** loaded for the language `#1'. Using the pattern for}%
\typeout{** the default language instead.}%
\else
\language=\csname l@#1\endcsname
\fi
#2}}
\providecommand{\BIBdecl}{\relax}
\BIBdecl

\bibitem{fu2008}
Y.~Fu and L.~Li, ``Generalized analytic signal associated with linear canonical
  transform,'' \emph{Optics Communications}, vol. 281, no.~6, pp. 1468--1472,
  2008.

\bibitem{gabor46}
D.~Gabor, ``Theory of communication. part 1: The analysis of information,''
  \emph{Journal of the Institution of Electrical Engineers-Part III: Radio and
  Communication Engineering}, vol.~93, no.~26, pp. 429--441, 1946.

\bibitem{feldman11}
M.~Feldman, \emph{Hilbert transform applications in mechanical
  vibration}.\hskip 1em plus 0.5em minus 0.4em\relax Wiley Online Library,
  2011.

\bibitem{taylor81}
L.~Taylor, ``The phase retrieval problem,'' \emph{IEEE Transactions on Antennas
  and Propagation}, vol.~29, no.~2, pp. 386--391, 1981.

\bibitem{boashash921}
B.~Boashash, ``Estimating and interpreting the instantaneous frequency of a
  signal. I. fundamentals,'' \emph{Proceedings of the IEEE}, vol.~80, no.~4,
  pp. 520--538, 1992.

\bibitem{boashash922}
B.~Boashash, ``Estimating and interpreting the instantaneous frequency of a signal.
  II. algorithms and applications,'' \emph{Proceedings of the IEEE}, vol.~80,
  no.~4, pp. 540--568, 1992.

\bibitem{marple1999}
S.~L. Marple, Jr., ``Estimating group delay and phase delay via discrete-time
  ``analytic'' cross-correlation,'' \emph{IEEE Transactions on Signal
  Processing}, vol.~47, no.~9, pp. 2604--2607, 1999.

\bibitem{boashash2003}
B.~Boashash, \emph{Time frequency signal analysis and processing: a
  comprehensive reference}.\hskip 1em plus 0.5em minus 0.4em\relax Elsevier
  Science, 2003.

\bibitem{huang1998}
N.~E. Huang, Z.~Shen, S.~R. Long, M.~C. Wu, H.~H. Shih, Q.~Zheng, N.~C. Yen,
  C.~C. Tung, and H.~H. Liu, ``The empirical mode decomposition and the Hilbert
  spectrum for nonlinear and non-stationary time series analysis,'' \emph{Proc.
  R. Soc. Lond. A}, vol. 454, no. 1971, pp. 903--995, 1998.

\bibitem{huang2005}
N.~E. Huang and S.~S. Shen, \emph{Hilbert-Huang transform and its
  applications}.\hskip 1em plus 0.5em minus 0.4em\relax World Scientific
  Publishing Co., 2005.

\bibitem{benitez2001}
D.~Benitez, P.~Gaydecki, A.~Zaidi, and A.~Fitzpatrick, ``The use of the Hilbert
  transform in ECG signal analysis,'' \emph{Computers in Biology and Medicine},
  vol.~31, no.~5, pp. 399--406, 2001.

\bibitem{wilson2008}
J.~D. Wilson, R.~B. Govindan, J.~O. Hatton, C.~L. Lowery, and H.~Preissl,
  ``Integrated approach for fetal QRS detection,'' \emph{IEEE Transactions on
  Biomedical Engineering}, vol.~55, no.~9, pp. 2190--2197, 2008.

\bibitem{collins1970}
S.~A. Collins, Jr., ``Lens-system diffraction integral written in terms of
  matrix optics,'' \emph{JOSA}, vol.~60, no.~9, pp. 1168--1177, 1970.

\bibitem{moshinsky1971}
M.~Moshinsky and C.~Quesne, ``Linear canonical transformations and their
  unitary representations,'' \emph{Journal of Mathematical Physics}, vol.~12,
  p. 1772, 1971.

\bibitem{wolf1979}
K.~B. Wolf, \emph{Integral transforms in science and engineering}.\hskip 1em
  plus 0.5em minus 0.4em\relax Plenum Press, 1979, ch.~9.

\bibitem{ozaktas2001}
H.~M. Ozaktas, M.~A. Kutay, and Z.~Zalevsky, \emph{The fractional Fourier
  transform with applications in optics and signal processing}.\hskip 1em plus
  0.5em minus 0.4em\relax New York: Wiley, 2001.

\bibitem{ding2001}
J.-J. Ding, ``Research of fractional Fourier transform and linear canonical
  transform,'' Ph.D. dissertation, Ph. D. Thesis, National Taiwan University,
  2001.

\bibitem{pei2002}
S.~C. Pei and J.-J. Ding, ``Eigenfunctions of linear canonical transform,''
  \emph{IEEE Transactions on Signal Processing}, vol.~50, no.~1, pp. 11--26,
  2002.

\bibitem{nazarathy1982}
M.~Nazarathy and J.~Shamir, ``First-order optics—a canonical operator
  representation: lossless systems,'' \emph{JOSA}, vol.~72, no.~3, pp.
  356--364, 1982.

\bibitem{bastiaans1989}
M.~J. Bastiaans, ``Propagation laws for the second-order moments of the Wigner
  distribution function in first-order optical systems,'' \emph{Optik},
  vol.~82, no.~4, pp. 173--181, 1989.

\bibitem{barshan1997}
B.~Barshan, M.~A. Kutay, and H.~M. Ozaktas, ``Optimal filtering with linear
  canonical transformations,'' \emph{Optics Communications}, vol. 135, no. 1-3,
  pp. 32--36, 1997.

\bibitem{pei2000}
S.~C. Pei and J.-J. Ding, ``Simplified fractional Fourier transforms,''
  \emph{JOSA A}, vol.~17, no.~12, pp. 2355--2367, 2000.

\bibitem{pei2001}
S.~C. Pei and J.-J. Ding, ``Relations between fractional operations and time-frequency
  distributions, and their applications,'' \emph{IEEE Transactions on Signal
  Processing}, vol.~49, no.~8, pp. 1638--1655, 2001.

\bibitem{bastiaans2003}
M.~J. Bastiaans and K.~B. Wolf, ``Phase reconstruction from intensity
  measurements in linear systems,'' \emph{JOSA A}, vol.~20, no.~6, pp.
  1046--1049, 2003.

\bibitem{hennelly2005}
B.~M. Hennelly and J.~T. Sheridan, ``Optical encryption and the space bandwidth
  product,'' \emph{Optics Communications}, vol. 247, no.~4, pp. 291--305, 2005.

\bibitem{sharma2006}
K.~K. Sharma and S.~D. Joshi, ``Signal separation using linear canonical and
  fractional Fourier transforms,'' \emph{Optics Communications}, vol. 265,
  no.~2, pp. 454--460, 2006.

\bibitem{lohmann1996}
A.~W. Lohmann, D.~Mendlovic, and Z.~Zalevsky, ``Fractional Hilbert transform,''
  \emph{Optics Letters}, vol.~21, no.~4, pp. 281--283, 1996.

\bibitem{pei2000b}
S.~C. Pei and M.-H. Yeh, ``Discrete fractional Hilbert transform,'' \emph{IEEE Transactions on Circuits and Systems II: Analog and Digital Signal Processing}, vol.~47, no.~11, pp. 1307--1311, 2000.


\bibitem{tseng2000}
C.-C. Tseng and S.~C. Pei, ``Design and application of discrete-time fractional
  Hilbert transformer,'' \emph{IEEE Transactions on Circuits and Systems II:
  Analog and Digital Signal Processing}, vol.~47, no.~12, pp. 1529--1533, 2000.

\bibitem{cusmariu2002}
A.~Cusmariu, ``Fractional analytic signals,'' \emph{Signal processing},
  vol.~82, no.~2, pp. 267--272, 2002.

\bibitem{tao2008}
R.~Tao, X.~M. Li, and Y.~Wang, ``Generalization of the fractional Hilbert
  transform,'' \emph{Signal Processing Letters, IEEE}, vol.~15, pp. 365--368,
  2008.

\bibitem{zayed1998}
A.~I. Zayed, ``Hilbert transform associated with the fractional Fourier
  transform,'' \emph{IEEE Signal Processing Letters}, vol.~5, no.~8, pp.
  206--208, 1998.

\bibitem{Gradshteyn2007}
I.~S. Gradshteyn and I.~M. Ryzhik, \emph{Table of integrals, series, and
  products, 7th ed.}\hskip 1em plus 0.5em minus 0.4em\relax San Diego, CA:
  Academic, 2007.

\bibitem{fadeeva1954}
V.~N. Fadeeva and N.~M. Terent'ev, ``Tables of values of the probability
  integral,'' \emph{Gostekhizdat, Moscow}, 1954.

\bibitem{hilbert1912}
D.~Hilbert, \emph{Grundz{\"u}ge einer allgemeinen Theorie der linearen
  Integralgleichungen}.\hskip 1em plus 0.5em minus 0.4em\relax Leipzig: B.~G.
  Teubner, 1912.

\bibitem{king2009}
F.~W. King, \emph{Hilbert transforms, Volumn 1}.\hskip 1em plus 0.5em minus
  0.4em\relax Cambridge University Press, 2009.

\bibitem{king2009v2}
F.~W. King, \emph{Hilbert transforms, Volumn 2}.\hskip 1em plus 0.5em minus
  0.4em\relax Cambridge University Press, 2009.

\bibitem{pei2011}
S.~C. Pei and Y.-C. Lai, ``Discrete linear canonical transforms based on
  dilated Hermite functions,'' \emph{JOSA A}, vol.~28, no.~8, pp. 1695--1708,
  2011.

\bibitem{lu2009}
W.~Lu and Q.~Zhang, ``Deconvolutive short-time Fourier transform spectrogram,''
  \emph{IEEE Signal Processing Letters}, vol.~16, no.~7, pp. 576--579, 2009.


\bibitem{pei2007}
S.~C. Pei and J.-J. Ding, ``Relations between Gabor transforms and fractional
  Fourier transforms and their applications for signal processing,'' \emph{IEEE
  Transactions on Signal Processing}, vol.~55, no.~10, pp. 4839--4850, 2007.

\end{thebibliography}
\bibliographystyle{IEEEtran}

\begin{IEEEbiography}[{\includegraphics[width=1in,height=1.25in,clip,keepaspectratio]{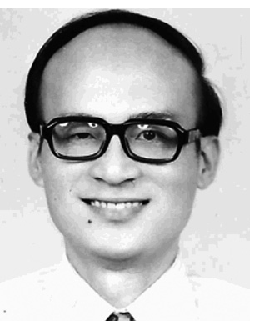}}]
{Soo-Chang Pei} (SM'89-F'00) was born in Soo-Auo, Taiwan, in 1949. He received the B.S.E.E. degree from National Taiwan University, Taipei, Taiwan, in 1970, and the M.S.E.E. and Ph.D. degrees from the University of California Santa Barbara, Santa Barbara, in 1972 and 1975, respectively. From 1970 to 1971, he was an Engineering Officer with the Chinese Navy Shipyard. From 1971 to 1975, he was a Research Assistant with the University of California Santa Barbara. He was a Professor and the Chairman of the Department of Electrical Engineering with the Tatung Institute of Technology, Taipei, from 1981 to 1983 and with National Taiwan University from 1995 to 1998. From 2003 to 2009, he was the Dean of the College of Electrical Engineering and Computer Science with National Taiwan University. He is currently a Professor with the Department of Electrical Engineering, National Taiwan University. His research interests include digital signal processing, image processing, optical information processing, and laser holography.

Dr. Pei was a recipient of the National Sun Yet-Sen Academic Achievement Award in Engineering in 1984, the Distinguished Research Award from the National Science Council from 1990 to 1998, the Outstanding Electrical Engineering Professor Award from the Chinese Institute of Electrical Engineering in 1998, the Academic Achievement Award in Engineering from the Ministry of Education in 1998, the Pan Wen-Yuan Distinguished Research Award in 2002, and the National Chair Professor Award from the Ministry of Education in 2002. He was the President of the Chinese Image Processing and Pattern Recognition Society in Taiwan from 1996 to 1998 and is a member of Eta Kappa Nu and the Optical Society of America. He became an IEEE Fellow in 2000 for his contributions to the development of digital eigenfilter design, color image coding and signal compression and to electrical engineering education in Taiwan.
\end{IEEEbiography}

\begin{IEEEbiography}[{\includegraphics[width=1in,height=1.25in,clip,keepaspectratio]{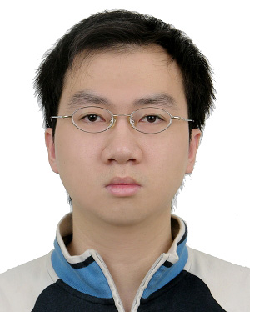}}]
{Shih-Gu Huang} received the B.S. degree in electrical engineering and the M.S. degree in communications engineering from National Tsing Hua University, Hsinchu, Taiwan, in 2007 and 2009, respectively. He is currently working toward the Ph.D. degree in the Graduate Institute of Communication Engineering, National Taiwan University, Taipei, Taiwan. His research interests include digital signal processing, time-frequency analysis, fractional Fourier transform, and linear canonical transform.
\end{IEEEbiography}

\vfill

\end{document}